\documentclass[11pt]{article}
\usepackage{amsmath} 
\usepackage{amscd}     \usepackage{amsxtra}
\usepackage{upref}     \usepackage{amsthm}
\usepackage{amssymb}   
\usepackage{amsfonts}
\begin{document}

\title{\bf
Frobenius Curvature, Electromagnetic Strain and Description of Photon-like
Objects}

\author{
{\bf Stoil Donev$^1$\footnote{e-mail:
 sdonev@inrne.bas.bg} \ , Maria Tashkova$^1$}
\noindent
\\ (1) Institute for Nuclear Research and Nuclear Energy,\\
Bulg.Acad.Sci., 1784 Sofia, blvd.Tzarigradsko chaussee 72, Bulgaria}
\date{}
\maketitle

\begin{abstract}
This paper aims to present a general idea for description of spatially finite physical objects with
a consistent nontrivial translational-rotational dynamical structure and evolution as a whole,
making use of the mathematical concepts and structures connected with the Frobenius
integrability/nonintegrability theorems and electromagnetic strain quantities. The idea is based on
consideration of {\it nonintegrable} subdistributions of some appropriate completely integrable
distribution (differential system) on a manifold and then to make use of the corresponding
curvatures and correspondingly directed strains as measures of interaction, i.e. of energy-momentum
exchange among the physical subsystems mathematically represented by the nonintegrable
subdistributions. The concept of photon-like object is introduced and description (including
lagrangian) of such objects in these terms is given.

 \end{abstract}

\section{Introduction} At the very dawn of the 20th century Planck (Planck 1901) proposed and a
little bit later Einstein (Einstien 1905) appropriately used the well known and widely used through
the whole last century simple formula $E=h\nu$, $h=const>0$.  This formula marked the beginning of
a new era and became a real symbol of the physical science during the following years. According to
the Einstein's interpretation it gives the full energy $E$ of {\it really existing} light quanta of
frequency $\nu=const$, and in this way a new understanding of the nature of the electromagnetic
field was introduced:  the field has structure which contradicts the description given by Maxwell
vacuum equations.  After De Broglie's (De Broglie 1923) suggestion for the particle-wave nature of
the electron obeying the same energy-frequency relation, one could read Planck's formula in the
following way:  {\it there are physical objects in Nature the very existence of which is strongly
connected to some periodic (with time period $T=1/\nu$) process of intrinsic for the object nature
and such that the Lorentz invariant product $ET$ is equal to $h$}. Such a reading should suggest
that these objects do NOT admit point-like approximation since the relativity principle for free
point particles requires straight-line uniform motion, hence, no periodicity should be allowed.

Although the great (from pragmatic point of view) achievements of the developed theoretical
approach, known as {\it quantum theory}, the great challenge to build an adequate description of
individual representatives of these objects, especially of light quanta called by Lewis {\it
photons} (Lewis 1926) is still to be appropriately met since  the efforts made in this direction,
we have to admit, still have not brought satisfactory results. Recall that Einstein in his late
years recognizes (Speziali 1972) that "the whole fifty years of conscious brooding have not brought
me nearer to the answer to the question "what are light quanta", and now, half a century later,
theoretical physics still needs progress to present a satisfactory answer to the question "what is
a photon". We consider the corresponding theoretically directed efforts as necessary and even {\it
urgent} in view of the growing amount of definite experimental skills in manipulation with
individual photons, in particular, in connection with the experimental advancement in the "quantum
computer" project.  The dominating modern theoretical view on microobjects is based on the notions
and concepts of quantum field theory (QFT) where the structure of the photon (as well as of any
other microobject) is accounted for mainly through the so called {\it structural function}, and
highly expensive and delicate collision experiments are planned and carried out namely in the frame
of these concepts and methods (see the 'PHOTON' Conferences Proceedings, some recent review papers:
Dainton 2000; Stumpf, Borne 2001; Godbole 2003; Nisius 2001).  Going not in details we just note a
special feature of this QFT approach: if the study of a microobject leads to conclusion that it has
structure, i.e. it is not point-like, then the corresponding constituents of this structure are
considered as point-like, so the point-likeness stays in the theory just in a lower level.

In this paper we follow another approach based on the assumption that the description of the
available (most probably NOT arbitrary) spatial structure of photon-like objects can be made by
{\it continuous finite/localized} functions of the three space variables.  The difficulties met in
this approach consist mainly, in our view, in finding adequate enough mathematical objects and
solving appropriate PDE.  The lack of sufficiently reliable corresponding information made us look
into the problem from as general as possible point of view on the basis of those properties of
photon-like objects which may be considered as most undoubtedly trustful, and
in some sense, {\it identifying}. The analysis made suggested that such a
property seems to be the fact that {\it the propagation of an individual
photon-like object necessarily includes a straight-line translational uniform
component}, so we shall focus on this property in order to see what useful for
our purpose suggestions could be deduced and what appropriate structures could
be constructed. All these suggestions and structures should be the building
material for a step-by-step creation a {\it self-consistent} system. From
physical point of view this should mean that the corresponding properties may combine to express a
dynamical harmony in the inter-existence of appropriately defined subsystems of
a finite and time stable larger physical system.

The plan of this paper is the following. In Sec.2 we introduce and comment the concept of {\it
photon-like object}. In Sec.3 we recall some basic facts from Frobenius integrability theory, then
we consider its possibilities to describe interaction between/among subsystems, mathematically
represented by non-integrable subdistributions of an integrable distribution, and finally we
introduce objects and structures in correspondence with the notion for photon-like object of Sec.2.
In Sec.4 we make use of these objects to define corresponding relativistic strain tensor(s) and
other related objects, and establish important relations with the curvature properties of the
subdistributions considered in Sec.3. In Sec.5 we show explicitly how the translational-rotational
consistency could be accounted for. In Sec.6 we consider possible lagrangian approaches giving
appropriate equations and spatially finite solutions with photon-like properties and behavior. In
the concluding Sec.6 we discuss the results obtained and present some views for further
development.

\section{The notion of photon-like object} We begin with the notice that any notion of a physical
object must unify two kinds of properties of the object considered: {\it identifying} or {\it
proper}, and {\it kinematical}. The identifying properties stay unchanged throughout the existence
of the object, and the kinematical properties describe those changes, called {\it admissible},
which do NOT lead to destruction of the object. Correspondingly, physics introduces two kinds of
quantities, proper and kinematical, but the more important quantities used in theoretical physics
turn out to be the {\it dynamical} quantities which, as a rule, are functions of the proper and
kinematical ones. In view of this we introduce the following notion of Photon-like object (we shall
use the abbreviation "PhLO" for "Photon-like object(s)"):

\begin{center} {\bf PhLO are real massless time-stable physical objects with a consistent
translational-rotational dynamical structure}. \end{center}

We give now some explanatory comments, beginning with the term {\it real}. {\bf First} we emphasize
that this term means that we consider PhLO as {\it really} existing {\it physical} objects, not as
appropriate and helpful but imaginary (theoretical) entities.  Accordingly, PhLO {\bf necessarily
carry energy-momentum}, otherwise, they could hardly be detected.  {\bf Second}, PhLO can
undoubtedly be {\it created} and {\it destroyed}, so, no point-like and infinite models are
reasonable: point-like objects are assumed to have no structure, so they can not be destroyed since
there is no available structure to be destroyed; creation of infinite objects (e.g. plane waves)
requires infinite quantity of energy to be transformed from one kind to another for finite
time-period, which seems also unreasonable. Accordingly, PhLO are {\it spatially finite} and have
to be modeled like such ones, which is the only possibility to be consistent with their
"created-destroyed" nature. It seems hardly believable that spatially infinite and indestructible
physical objects may exist at all.  {\bf Third}, "spatially finite" implies that PhLO may carry
only {\it finite values} of physical (conservative or non-conservative) quantities.  In particular,
the most universal physical quantity seems to be the energy-momentum, so the model must allow
finite integral values of energy-momentum to be carried by the corresponding solutions. {\bf
Fourth}, "spatially finite" means also that PhLO {\it propagate}, i.e.  they do not "move" like
classical particles along trajectories, therefore, partial differential equations should be used to
describe their evolution in time.

The term "{\bf massless}" characterizes physically the way of propagation: the {\it integral}
energy $E$ and {\it integral} momentum $p$ of a PhLO should satisfy the relation $E=cp$, where $c$
is the speed of light in vacuum, and in relativistic terms this means that their integral
energy-momentum vector {\it must be isotropic}, i.e. it must have zero module with respect to
Lorentz-Minkowski (pseudo)metric in $\mathbb{R}^4$. If the object considered has spatial and
time-stable structure, so that the translational velocity of every point where the corresponding
field functions are different from zero must be equal to $c$, we have in fact null direction in the
space-time intrinsically determined by the PhLO. Such a direction is formally
defined by a null vector field $\bar{\zeta},\bar{\zeta}^2=0$. The integral
trajectories of this vector field are isotropic (or null) straight lines. It
follows that with every PhLO a null direction is {\it necessarily} associated,
so, canonical coordinates $(x^1,x^2,x^3,x^4)=(x,y,z,\xi=ct)$ on $\mathbb{R}^4$
may be chosen such that in the corresponding coordinate frame $\bar{\zeta}$ to
have only two non-zero components of magnitude $1$:
$\bar{\zeta}^\mu=(0,0,-\varepsilon, 1)$, where $\varepsilon=\pm 1$ accounts for
the two directions along the coordinate $z$. Further such a coordinate system
will be called $\bar{zeta}$-adapted and will be of main usage. It may be also
expectable, that the corresponding energy-momentum tensor $T_{\mu\nu}$ of the model satisfies the
relation $T_{\mu\nu}T^{\mu\nu}=0$, which may be considered as a localization of the integral
isotropy condition $E^2-c^2p^2=0$.

The term "{\bf translational-rotational}" means that besides translational component along
$\bar{\zeta}$, the propagation necessarily demonstrates some rotational (in the general sense of
this concept) component in such a way that {\it both components exist simultaneously and
consistently}. It seems reasonable to expect that such kind of behavior should be consistent only
with some distinguished spatial shapes. Moreover, if the Planck relation $E=h\nu$ must be respected
throughout the evolution, the rotational component of propagation should have {\it time-periodical}
nature with time period $T=E/\nu=const$, and one of the two possible, {\it left} or {\it right},
orientations. It seems reasonable also to expect periodicity in the spatial shape of PhLO, which
somehow to be related to the time periodicity.

The term "{\bf dynamical structure}" means that the propagation is supposed to be necessarily
accompanied by an {\it internal energy-momentum redistribution}, which may be considered in the
model as energy-momentum exchange between (or among) some appropriately defined subsystems.  It
could also mean that PhLO live in a dynamical harmony with the outside world, i.e.  {\it any
outside directed energy-momentum flow should be accompanied by a parallel inside directed
energy-momentum flow}.

Finally, note that if the time periodicity and the spatial periodicity
should be consistently related somehow, the simplest such consistency
would seem like this: the spatial size along the translational component
of propagation $4l_o$ is equal to $cT$: $4l_o=cT$, where $l_o$ is some
finite positive characteristic constant of the corresponding solution.
This would mean that every individual PhLO determines its own length/time
scale.
\vskip0.3cm
We are going now to formulate shortly the basic idea inside which this study will be carried out.

\section{Frobenius curvature and interaction}

Any physical system with a dynamical structure is characterized with some
internal energy-momentum redistributions, i.e. energy-momentum fluxes, during
evolution. Any system of energy-momentum fluxes (as well as fluxes of other
interesting for the case physical quantities subject to change during
evolution, but we limit ourselves just to energy-momentum fluxes here) can be
considered mathematically as generated by some system of vector fields. A {\it
consistent} and {\it interelated time-stable} system of energy-momentum fluxes
can be considered to correspond directly or indirectly to an integrable
distribution $\Delta$ of vector fields according to the principle {\it local
object generates integral object}. It seems reasonable to assume the following
geometrization of the concept of physical interaction: {\it two distributions
$\Delta_1$ and $\Delta_2$ on a manifold will be said to interact geometrically
if at least one of the corresponding two curvature forms $\Omega_1$/$\Omega_2$
takes values, or generates objects taking values, respectively in
$\Delta_2$/$\Delta_1$}.

The above concept of {\it geometrical interaction} is motivated by the fact
that, in general, an integrable distribution $\Delta$ may contain various {\it
nonintegrable} subdistributions $\Delta_1, \Delta_2, \dots$ which
subdistributions may be interpreted physically as interacting subsytems. Any
physical interaction between 2 subsystems is necessarily accompanied with
available energy-momentum exchange between them, this could be understood
mathematically as nonintegrability of each of the two subdistributions of
$\Delta$ and could be naturally measured directly or indirectly by the
corresponding curvatures. For example, if $\Delta$ is an integrable
3-dimensional distribution spent by the vector fields $(X_1,X_2,X_3)$ then we
may have, in general, three non-integrable, i.e. geometrically interacting,
2-dimensional subdistributions $(X_1,X_2), (X_1,X_3), (X_2,X_3)$. Finally, some
interaction with the outside world can be described by curvatures of
distributions (and their subdistributions) in which elements from $\Delta$ and
vector fields outside $\Delta$ are involved (such processes will not be
considered in this paper).

To make the above statements mathematically clearer we recall the Frobenius theorem on a manifold
$M^n$ [Godbillon 1969] (further all manifolds are assumed smooth and finite dimensional and all
objects defined on $M^n$ are also assumed smooth). If the system of vector
fields $\Delta=\left[X_1(x),X_2(x),\dots, X_p(x)\right]$, $x\in M$, $1<p<n$,
satisfies $X_1(x)\wedge X_2(x)\wedge \dots ,\wedge\, X_p(x)\neq 0, \,x\in M$
then $\Delta$ is integrable iff all Lie brackets $\left[X_i,X_j\right], \
i,j=1,2,\dots, p$ are representable linearly through the very $X_i,
i=1,2,\dots, p: \left[X_i,X_j\right]=C^k_{ij}X_k$, where $C^k_{ij}$ are
functions. Clearly, an easy way to find out if a distribution is integrable is
to check if the exterior products $$ [X_i,X_j]\wedge X_1(x)\wedge X_2(x)\wedge
\dots ,\wedge\, X_p(x), \,x\in M;\ \ \ i,j=1,2,\dots,p $$ are identically zero.
If this is not the case (which means that at least one such Lie bracket "sticks
out" of the distribution $\Delta$) then the corresponding coefficients, which
are multilinear combinations of the components of the vector fields
and their derivatives, represent the corresponding curvatures. We note
finally that if two subdistributions contain at least one common vector field
it seems naturally to expect interaction.

In the dual formulation of Frobenius theorem in terms of differential 1-forms (i.e. Pfaff forms) we
look for $(n-p)$-Pfaff forms $(\alpha^1, \alpha^2, \dots, \alpha^{n-p}$), i.e. a
$(n-p)$-codistribution $\Delta^*$, such that $ \langle\alpha^m,X_j\rangle=0,\
\ \text{and} \ \ \alpha^1\wedge\alpha^2\wedge\dots \wedge\alpha^{n-p}\neq 0, $
$ m=1,2,\dots,n-p, \ \ j=1,2,\dots,p . $ Then the integrability of the
distribution $\Delta$ is equivalent to the requirements $$
\mathbf{d}\alpha^m\wedge\alpha^1\wedge\alpha^2\wedge\dots\wedge\alpha^{n-p} =0,\ \ \ m=1,2,\dots,
(n-p), $$ where $\mathbf{d}$ is the exterior derivative.

Since the idea of curvature associated with, for example, an arbitrary
2-dimensional
distribution $(X,Y)$ is to find out if the Lie bracket $[X,Y]$ has components
along vector fields outside the 2-plane defined by $(X,Y)$, in our case
 we have to evaluate
the quantities $\langle\alpha^m,[X,Y]\rangle$, where all linearly independent
1-forms $\alpha^m$ annihilate
$(X,Y):\langle\alpha^m,X\rangle=\langle\alpha^m,Y\rangle=0$. In view of the
formula
$$ \mathbf{d}\alpha^m(X,Y)=X(\langle\alpha^m,Y\rangle)-
Y(\langle\alpha^m,X\rangle) -\langle\alpha^m,[X,Y]\rangle=
-\langle\alpha^m,[X,Y]\rangle
$$
we may introduce explicitly the curvature 2-form for the distribution
$\Delta(X)=(X_1,\dots,X_p)$. In
fact, if $\Delta(Y)=(Y_1,\dots,Y_{n-p})$ define a distribution which is
complimentary to $\Delta(X)$ and
$\langle\alpha^m,Y_n\rangle=\delta^m_n$, i.e. $(Y_1,\dots,Y_{n-p})$ and
$(\alpha^1, \dots, \alpha^{n-p})$ are dual bases, then the
corresponding curvature 2-form $\Omega_{\Delta(X)}$ should be defined by
$$
\Omega_{\Delta(X)}=-\mathbf{d}\alpha^m\otimes Y_m, \ \ \text{since} \ \
\Omega_{\Delta(X)}(X_i,X_j)=-\mathbf{d}\alpha^m(X_i,X_j) Y_m=
\langle\alpha^m,[X_i,X_j]\rangle Y_m ,
$$
where it is meant here that
$\Omega_{\Delta(X)}$ is restricted to the distribution $(X_1,\dots,X_p)$.
Hence, if we call the distribution $(X_1,\dots,X_p)$ {\it horizontal}, and the
complimentary distribution $(Y_1,\dots,Y_{n-p})$ {\it vertical} then the
curvature 2-form acquires the status of {\it vertical bundle valued 2-form}. We
see that the curvature 2-form distinguishes those couples of vector fields
inside $\Delta(X)$ the Lie brackets of which define outside $\Delta(X)$
directed flows, and so, do not allowing to find integral manifold of
$\Delta(X)$. Clearly, the supposition here for dimensional complementarity of
the two distributions $\Delta(X)$ and $\Delta(Y)$ is not essential for the idea
of curvature.

Hence, from physical point of view, if we make use of the quantities
$\Omega_{\Delta(X)}(X_i,X_j)$ to build the
components of the energy-momentum locally transferred from the system
$\Delta(X)$ to the system $\Delta(Y)$, then, naturally, we have to make use of
the quantities
$\Omega_{\Delta(Y)}(Y_m,Y_n)$ to build the components of
the energy-momentum transferred from $\Delta(Y)$ to $\Delta(X)$.
It deserves to note that it is possible a dynamical
equilibrium between the two systems $\Delta(Y)$ and
$\Delta(X)$ to exist: each system to gain as much energy-momentum as it
loses, and this to take place at every space-time point. On the other hand, the
restriction of $\Omega_{\Delta(X)}=-\mathbf{d}\alpha^m\otimes Y_m,
m=1,\dots,n-p,$ to the system $\Delta(Y)$, i.e. the quantities
$\Omega_{\Delta(X)}(Y_m,Y_n)$, and the restriction of
$\Omega_{\Delta(Y)}=-\mathbf{d}\beta^i\otimes X_i, i=1,\dots,p,
\langle\beta^i, X_j\rangle=\delta^i_j,
\beta^1\wedge\dots\wedge\beta^p\neq 0,
\langle\beta^m,Y_i\rangle=0$, to $\Delta(X)$, i.e. the quantities
$\Omega_{\Delta(Y)}(X_i,X_j)$, acquire the sense of objects causing local
change of the corresponding energy-momentum, i.e. differences between
energy-momentum gains and losses. Therefore, if $W_{(X,Y)}$ denotes the
energy-momentum transferred from $\Delta(X)$ to $\Delta(Y)$,  $W_{(Y,X)}$
denotes the energy-momentum transferred from $\Delta(Y)$ to $\Delta(X)$, and
$\delta W_{(X)}$ and $\delta W_{(Y)}$ denote respectively the energy-momentum
changes of the two systems $\Delta(X)$ and $\Delta(Y)$, then according to the
local energy-momentum conservation law we can write
\[
\delta W_{(X)}=W_{(Y,X)}+W_{(X,Y)}, \ \ \delta
W_{(Y)}=-(W_{(X,Y)}+W_{(Y,X)})=-\delta W_{(X)} .
\]
For the case of dynamical
equilibrium we have $\delta W_{(X)}=\delta W_{(Y)}=0$, so we obtain \[ \delta
W_{(X)}=0,\ \ \ \delta W_{(Y)}=0,\ \ \ W_{(Y,X)}+W_{(X,Y)}=0. \] As for how to
build explicitly the corresponding representatives of the energy-momentum
fluxes, probably, universal procedure can not be offered since the adequate
mathematical representative of the system under consideration depends strongly
on the very system. If, for example, the mathematical representative is a
differential form $G$, then the most simple procedure seems to be to "project"
the curvature components $\Omega_{\Delta(X)}(X_i,X_j)$ and
$\Omega_{\Delta(Y)}(Y_m,Y_n)$, as well as the components
$\Omega_{\Delta(X)}(Y_i,Y_j)$ and $\Omega_{\Delta(Y)}(X_m,X_n)$ on $G$ i.e. to
consider the corresponding interior products. In the general case, appropriate
quantities constructed out of the members of the introduced distributions and
codistributions must be found.

Finally we note that, as we shall see further, a PhLO may be considered to
represent an example of a system, functioning through a dynamical equilibrium
between two appropriately defined and interacting subsystems.
\vskip 0.2cm
We are going now to make use of the above general consideration to find
appropriate objects and relations  in an attempt to describe PhLO's dynamical
structure and evolution in these terms.

\section{PhLO dynamical structure in terms of Frobenius curvature}
We consider the Minkowski space-time $M=(\mathbb{R}^4,\eta)$ with
signature $sign(\eta)=(-,-,-,+)$ related to the standard global
coordinates $$ (x^1,x^2,x^3,x^4)=(x,y,z,\xi=ct), $$ and the natural volume
form
$\omega_o=\sqrt{|\eta|}dx^1\wedge dx^2\wedge dx^3\wedge dx^4=
dx\wedge dy\wedge dz\wedge d\xi$.

We introduce the null vector field $\bar{\zeta},\ \bar{\zeta}^2=0$, which in the
$\bar{\zeta}$-adapted coordinates (throughout used further) is assumed to look
as follows:
\begin{equation}
\bar{\zeta}=-\varepsilon\frac{\partial}{\partial z} +          
\frac{\partial}{\partial \xi}, \ \ \varepsilon=\pm 1.
\end{equation}
Let's denote the corresponding to $\bar{\zeta}$ completely integrable
3-dimensional Pfaff system by $\Delta^*(\bar{\zeta})$. Thus,
$\Delta^*(\bar{\zeta})$ is generated by three linearly independent 1-forms
$(\alpha_1,\alpha_2,\alpha_3)$ which annihilate $\bar{\zeta}$, i.e. $$
\alpha_1(\bar{\zeta})=\alpha_2(\bar{\zeta})=\alpha_3(\bar{\zeta})=0; \ \
\alpha_1\wedge \alpha_2\wedge \alpha_3\neq 0.
$$
Instead of $(\alpha_1,\alpha_2,\alpha_3)$ we introduce the notation
$(A, A^*, \zeta)$ and define $\zeta$ by
\begin{equation}
\zeta=\varepsilon dz+d\xi,      
\end{equation}
Now, since $\zeta$ defines 1-dimensional completely integrable Pfaff
system we have the corresponding completely integrable distribution
$(\bar{A},\bar{A^{*}},\bar{\zeta})$. We shall restrict our further study on
PhLO of electromagnetic nature according to the following
\vskip 0.3cm
{\bf{Definition}}: We shall call a PhLO {\it electromagnetic} if it satisfies the following
conditions ($\langle , \rangle$ is the coupling between forms and
vectors):

1. $\langle\zeta,\bar{A}\rangle=\langle\zeta,\bar{A^{*}}\rangle=0$,

2. the vector fields $(\bar{A},\bar{A^{*}})$ have no components along
$\bar{\zeta}$,

3. $(\bar{A},\bar{A^{*}})$ are $\eta$-corresponding to $(A, A^*)$
respectively .

4. $\langle A,\bar{A^{*}}\rangle=0,\ \
\langle A,\bar{A}\rangle =\langle A^{*},\bar{A^{*}}\rangle$ .

\vskip 0.3cm
\noindent
{\bf Remark}. These relations formalize knowledge from Classical
electrodynamics. In fact, our vector fields $(\bar{A},\bar{A^{*}})$ are
meant to represent the electric $\mathbf{E}$ and magnetic $\mathbf{B}$
components of a free time-dependent electromagnetic field, where, as is well
known [Synge,1958], the translational propagation of the field energy-momentum
along a fixed null direction with the velocity "$c$" is possible only if the
two invariants $I_1=\mathbf{B}^2-\mathbf{E}^2$ and $I_2=2\mathbf{E}.\mathbf{B}$
are zero, because only in such a case the energy-momentum tensor has {\it
unique} null eigen direction. So it seems naturally to consider this property as
{\it intrinsic} for the field and to choose it as a starting point. Moreover,
in such a case the relation $(I_1)^2+(I_2)^2=0$ is equivalent to
$\mathbf{E}^2+\mathbf{B}^2=2|\mathbf{E}\times\mathbf{B}|$ and this relation
shows that this is the only case when a nonzero field momentum can
not be made equal to zero by means of frame change. Together with the fact
that the spatial direction of translational energy-momentum propagation is
determined by $\mathbf{E}\times\mathbf{B}$, this motivates to introduce the
vector field $\bar{\zeta}$ in this form and to assume the properties 1-4 in the
above definition. \vskip 0.3cm
 From the above conditions it follows that in the
$\bar{\zeta}$-adapted coordinate system we have
\[
A=u\,dx + p\,dy, \ \
A^*=-\varepsilon\,p\,dx + \varepsilon\,u\,dy; \ \
\bar{A}=-u\,\frac{\partial}{\partial x} -          
p\,\frac{\partial}{\partial y}, \ \
\bar{A^*}=\varepsilon\,p\,\frac{\partial}{\partial x} -            
\varepsilon\,u\,\frac{\partial}{\partial y},
\]
where $\varepsilon=\pm 1$, and $(u,p)$ are two smooth functions on $M$.

The completely integrable 3-dimensional Pfaff system $(A, A^*, \zeta)$
contains three 2-dimensional subsystems: $(A,A^{*}), (A,\zeta)$ and
$(A^*,\zeta)$. We have the following \vskip 0.3cm {\bf Proposition 1}. The
following relations hold:
\[
\mathbf{d}A\wedge A\wedge A^*=0;\ \
\mathbf{d}A^*\wedge A^*\wedge A=0;\ \                              
\]
\[ \mathbf{d}A\wedge A\wedge \zeta= \varepsilon\big[u(p_\xi-\varepsilon
p_z)- p(u_\xi-\varepsilon u_z)\big]\omega_o; \] \[ \mathbf{d}A^*\wedge
A^*\wedge \zeta=                                  
\varepsilon\big[u(p_\xi-\varepsilon p_z)- p(u_\xi-\varepsilon
u_z)\big]\omega_o.
\]
{\bf Proof.} Immediately verified.
\vskip 0.3cm
\noindent
These relations say that the 2-dimensional Pfaff system
$(A,A^*)$ is completely integrable for any choice of the two functions
$(u,p)$, while the two 2-dimensional Pfaff systems $(A,\zeta)$ and
$(A^*,\zeta)$ are NOT completely integrable in general, and the same
curvature factor $$ \mathbf{R}=u(p_\xi-\varepsilon
p_z)-p(u_\xi-\varepsilon u_z) $$ determines their nonintegrability.

Correspondingly, the 3-dimensional completely
integrable distribution (or differential system) $\Delta(\bar{\zeta})$
contains three 2-dimensional subsystems:
$(\bar{A},\bar{A^*})$, $(\bar{A},\bar{\zeta})$ and $(\bar{A^*},\bar{\zeta})$.
We have the
\vskip 0.3cm
{\bf Proposition 2}. The following relations hold (recall that $[X,Y]$ denotes the Lie
bracket):
\begin{equation}
[\bar{A},\bar{A^*}]\wedge\bar{A}\wedge\bar{A^*}=0 ,
\end{equation}
\begin{equation}
[\bar{A},\bar{\zeta}]=(u_\xi-\varepsilon u_z)\frac{\partial}{\partial x}+
(p_\xi-\varepsilon p_z)\frac{\partial}{\partial y} ,            
\end{equation}
\begin{equation}
[\bar{A^*},\bar{\zeta}]=
-\varepsilon(p_\xi-\varepsilon p_z)\frac{\partial}{\partial x}+
\varepsilon(u_\xi-\varepsilon u_z)\frac{\partial}{\partial y}.            
\end{equation}

{\bf Proof.} Immediately verified.
\vskip 0.3cm

From these last relations (3)-(5)
it follows that the distribution $(\bar{A},\bar{A^*})$
is integrable, and it can be easily shown that the
two distributions $(\bar{A},\bar{\zeta})$
and
$(\bar{A^*},\bar{\zeta})$ would be completely integrable only if the same
curvature factor
\begin{equation}
\mathbf{R}=u(p_\xi-\varepsilon p_z)-p(u_\xi-\varepsilon u_z)     
\end{equation}
is zero (the elementary proof is omitted).

As it should be, the two projections
$$
\langle A,[\bar{A^*},\bar{\zeta}]\rangle=
-\langle A^*,[\bar{A},\bar{\zeta}]\rangle=
\varepsilon u(p_\xi-\varepsilon p_z)-\varepsilon p(u_\xi-\varepsilon u_z)=
-\varepsilon\,\mathbf{R}
$$
are nonzero and give (up to a sign) the same factor $\mathbf{R}$. The same curvature factor
appears, of course, as coefficient in the exterior products
$
[\bar{A^*},\bar{\zeta}]\wedge \bar{A^*}\wedge\bar{\zeta}$ and
$[\bar{A},\bar{\zeta}]\wedge \bar{A}\wedge\bar{\zeta}$.
In fact, we obtain
\[
[\bar{A^*},\bar{\zeta}]\wedge \bar{A^*}\wedge\bar{\zeta}=
-[\bar{A},\bar{\zeta}]\wedge \bar{A}\wedge\bar{\zeta}=
-\varepsilon\mathbf{R}\,\frac{\partial}{\partial x}
\wedge\frac{\partial}{\partial y}\wedge
\frac{\partial}{\partial z}+
\mathbf{R}\,\frac{\partial}{\partial x}\wedge\frac{\partial}{\partial y}\wedge
\frac{\partial}{\partial \xi} .
\]
On the other hand, for the other two projections we obtain
\begin{equation}
\langle
A,[\bar{A},\bar{\zeta}]\rangle=                          
\langle
A^*,[\bar{A^*},\bar{\zeta}]\rangle= \frac12\big[(u^2+p^2)_\xi-\varepsilon
(u^2+p^2)_z\big].
\end{equation}
Clearly, the last relation (7) may be put in
terms of the Lie derivative $L_{\bar{\zeta}}$ as
$$
\frac 12L_{\bar{\zeta}}(u^2+p^2)=
-\frac12L_{\bar{\zeta}}\langle A,\bar{A}\rangle=
-\langle A,L_{\bar{\zeta}}\bar{A}\rangle=
-\langle A^*,L_{\bar{\zeta}}\bar{A^*}\rangle.
$$

{\bf Remark}. Further in the paper we shall denote $\sqrt{u^2+p^2}\equiv
\phi$, and shall assume that $\phi$ is a {\it spatially finite} function,
so, $u$ and $p$ must also be spatially finite. \vskip 0.3cm {\bf
Proposition 3.} There is a function $\psi(u,p)$ such, that

$$L_{\bar{\zeta}}\psi=\frac
{u(p_\xi-\varepsilon p_z)-p(u_\xi-\varepsilon u_z)}{\phi^2}=
\frac{\mathbf{R}}{\phi^2} .
$$
\vskip 0.4cm
{\bf Proof}. It is immediately verified that $\psi=\arctan\frac pu$
is such one.
\vskip 0.4cm
We note that the function $\psi$ has a natural
interpretation of {\it phase} because of the easily verified now relations
$u=\phi\cos\psi$, $p=\phi \sin\psi $,
and $\phi$ acquires the status of {\it amplitude}. Since the transformation
$(u,p)\rightarrow (\phi,\psi)$ is non-degenerate this allows to
work with the two functions $(\phi,\psi)$ instead of $(u,p)$.

From {\bf Prop.3} we have
\begin{equation}
\mathbf{R}=\phi^2L_{\bar{\zeta}}\psi= \
\phi^2(\psi_\xi-\varepsilon\psi_z) \,.             
\end{equation}

This last formula (8) shows something very important: at any $\phi\ne 0$
the curvature $\mathbf{R}$ will NOT be zero only if
$L_{\bar{\zeta}}\psi\neq 0$, which admits in principle availability of
rotation. In fact, lack of rotation would mean that $\phi$ and $\psi$ are
running plane waves along $\bar{\zeta}$. The relation $L_{\bar{\zeta}}\psi\neq
0$ means, however, that rotational properties are possible in general, and
some of these properties are carried by the phase $\psi$. It follows that
in such a case the translational component of propagation along
$\bar{\zeta}$ (which is supposed to be available) must be determined
essentially, and most probably entirely, by $\phi$.  In particular, we
could expect the relation $L_{\bar{\zeta}}\phi=0$ to hold, and if this
happens, then the rotational component of propagation will be represented
entirely by the phase $\psi$, and, more specially, by the curvature factor
$\mathbf{R}\neq 0$, so, since the objects we are going to describe have
consistent translational-rotational dynamical structure, further we assume
that, in general, $L_{\bar{\zeta}}\psi\neq 0$.

We are going now to represent some relations, analogical to the energy-momentum
relations in classical electrodynamics, determined by some 2-form $F$, in
terms of the Frobenius curvatures given above.

The two nonintegrable Pfaff systems $(A,\zeta)$ and
$(A^*,\zeta)$ define naturally corresponding 2-forms:
$$
G=A\wedge\zeta \ \
\text{and} \ \ G^*=A^*\wedge\zeta .
$$
We have also the 2-vectors
$$
\bar{G}=\bar{A}\wedge\bar{\zeta}, \ \ \text{and} \ \ \
\bar{G^*}=\bar{A^*}\wedge\bar{\zeta} .
$$
Making use now of the Hodge $*$-operator, defined by $\eta$ through the
relation $\alpha\wedge\beta=\eta(*\alpha,\beta)\omega_o$, where $\alpha$
and $\beta$ are $p$ and $(4-p)$-forms on $M$, we can verify the relation:
$G^*=*G$.
The 2-forms $G$ and $G^*$ define the 2-tensor, called
stress-energy-momentum tensor $T_{\mu}^{\nu}$, according to the rule
$$
T_{\mu}^{\nu}=-\frac12\big[G_{\mu\sigma}G^{\nu\sigma}+
(G^*)_{\mu\sigma}(G^*)^{\nu\sigma}\big] ,
$$
and the divergence of this tensor field can be represented in the form
$$
\nabla_\nu T_{\mu}^{\nu}=\big[i(\bar{G})\mathbf{d}G\big]_{\mu}+
\big[i(\bar{G^*})\mathbf{d}G^*\big]_\mu,
$$
where $\bar{G}$ and $\bar{G^*}$ coincide with the metric-corresponding
contravarint tensor fields, and $i(\bar{G})=i(\bar{\zeta})\circ i(\bar{A})$,
$i(\bar{G^*})=i(\bar{\zeta})\circ i(\bar{A^*})$,
$i(X)$ is the standard insertion operator in the exterior
algebra of differential forms on $\mathbb{R}^4$ defined by the vector field
$X$.
So, we shall need the quantities
$$
i(\bar{G})\mathbf{d}G, \ \ i(\bar{G^*})\mathbf{d}G^*,
\ \
i(\bar{G^*})\mathbf{d}G, \ \ i(\bar{G})\mathbf{d}G^* .
$$
Having in view the
explicit expressions for $A,A^*,\zeta,\bar{A},\bar{A^*}$ and $\bar{\zeta}$ we
obtain
\begin{equation}
i(\bar{G})\mathbf{d}G=i(\bar{G}^*)\mathbf{d}G^*=
\frac12 L_{\bar{\zeta}}\left(\phi^2\right).
\,\zeta \ .
\end{equation}
Also, we obtain
\[
i(\bar{G^*})\mathbf{d}G=-i(\bar{G})\mathbf{d}G^* =
\]
\begin{equation}
=\Big[u(p_\xi-\varepsilon p_z)-p(u_\xi-\varepsilon u_z)\Big]dz+
\varepsilon\Big[u(p_\xi-\varepsilon p_z)-p(u_\xi-\varepsilon u_z)\Big]d\xi
=\varepsilon\mathbf{R}\,\zeta .
\end{equation}
In the following formulae we must keep in mind
the relations
$\mathbf{d}\zeta=0, \langle A,\bar{A^*}\rangle=\langle\zeta,\bar{A^*}\rangle=
\langle\zeta,\bar{A}\rangle=0$.

The two distributions $(\bar{A},\bar{\zeta})$ and $(\bar{A^*},\bar{\zeta})$
determine corresponding curvature forms $\Omega$ and $\Omega^*$ according to
$$
\Omega=-\frac{1}{\phi^2}\mathbf{d}A^*\otimes\bar{A^*} ,\ \
\Omega^*=-\frac{1}{\phi^2}\mathbf{d}A\otimes\bar{A}.
$$
Denoting $Z_{\Omega}\equiv\Omega(\bar{A},\bar{\zeta})$,
$Z^*_{\Omega}\equiv\Omega(\bar{A^*},\bar{\zeta})$,
$Z_{\Omega^*}\equiv\Omega^*(\bar{A},\bar{\zeta})$ and
$Z^*_{\Omega^*}\equiv\Omega^*(\bar{A^*},\bar{\zeta})$
we obtain
\begin{equation}
Z_{\Omega}=-\frac{\varepsilon\mathbf{R}}{\phi^2}\bar{A^*},\ \
Z^*_{\Omega}=\frac{\bar{A^*}}{2\phi^2}L_{\bar{\zeta}}(\phi^2), \ \
Z_{\Omega^*}=\frac{\bar{A}}{2\phi^2}L_{\bar{\zeta}}(\phi^2), \ \
Z^*_{\Omega^*}=-\frac{\varepsilon\mathbf{R}}{\phi^2}\bar{A}.
\end{equation}
The following relations express the connection between the curvatures and the
energy-momentum characteristics.
\begin{align}
i(Z_{\Omega})(A\wedge\zeta)=0, \ \
i(Z_{\Omega})(A^*\wedge\zeta)=\varepsilon\mathbf{R}.\zeta=
-i(\bar{G})\mathbf{d}G^*=i(\bar{G^*})\mathbf{d}G, \\
i(Z_{\Omega^*})(A^*\wedge\zeta)=0, \ \
i(Z_{\Omega^*})(A\wedge\zeta)=\varepsilon\mathbf{R}.\zeta=
-i(\bar{G})\mathbf{d}G^*=i(\bar{G^*})\mathbf{d}G, \\
i(Z^*_{\Omega})(A\wedge\zeta)=0, \ \
i(Z^*_{\Omega})(A^*\wedge\zeta)=-\frac12L_{\bar{\zeta}}(\phi^2).\zeta=
-i(\bar{G})\mathbf{d}G=-i(\bar{G^*})\mathbf{d}G^{*}, \\
i(Z^*_{\Omega^*})(A^*\wedge\zeta)=0, \ \
i(Z^*_{\Omega^*})(A\wedge\zeta)=-\frac12L_{\bar{\zeta}}(\phi^2).\zeta=
-i(\bar{G})\mathbf{d}G=-i(\bar{G^*})\mathbf{d}G^{*}.
\end{align}

It follows from these relations that in case of dynamical equilibrium we shall
have
\[
i(\bar{G})\mathbf{d}G=0,\ \ i(\bar{G^*})\mathbf{d}G^*=0, \ \
i(\bar{G^*})\mathbf{d}G+i(\bar{G})\mathbf{d}G^*=0.
\]

Resuming, we can say that Frobenius integrability viewpoint suggests to
make use of one completely integrable 3-dimensional distribution (resp.
Pfaff system) consisting of one isotropic and two space-like vector fields
(resp. 1-forms), such that the corresponding 2-dimensional spatial
subdistribution $(\bar{A},\bar{A^*})$ (resp. Pfaff system $(A,A^*)$)
defines a completely integrable system, and the rest two 2-dimensional
subdistributions $(\bar{A},\bar{\zeta})$ and $(\bar{A^*},\bar{\zeta})$
(resp. Pfaff systems $(A,\zeta )$ and $(A^*,\zeta )$) are NON-integrable
in general and give the same curvature. This curvature may be used to
build quantities, physically interpreted as energy-momentum
internal exchanges between the corresponding
two subsystems $(\bar{A},\bar{\zeta})$ and $(\bar{A^*},\bar{\zeta})$
(resp.$(A,\zeta)$ and $(A^*,\zeta))$. Moreover, rotational component of
propagation will be available only if the curvature $\mathbf{R}$ is nonzero,
i.e. only if an internal energy-momentum exchange takes place. We see that
all physically important characteristics and relations, describing the
translational and rotational components of propagation, can be expressed in
terms of the corresponding Frobenius curvature. We'll see that this
holds also for some integral characteristics of PhLO.

\section{The electromagnetic strain viewpoint}
The concept of {\it strain}
is introduced in studying elastic materials subject to external forces of
different nature: mechanical, electromagnetic, etc. In nonrelativistic
continuum physics the local representatives of the external forces in this
context are usually called {\it stresses}. Since the force means
energy-momentum transfer leading to corresponding mutual energy-momentum
change of the interacting objects, then according to the energy-momentum
conservation law the material must react somehow to the external
interference in accordance with its structure and reaction abilities. The
classical strain describes mainly the abilities of the material to bear
force-action from outside through deformation, i.e. through changing its
shape, or, configuration. The term {\it elastic} now means that any two
allowed configurations can be deformed to each other without appearence of
holes and breakings, in particular, if the material considered has
deformed from configuration $C_1$ to configuration $C_2$ it is able to
return smoothly to its configuration $C_1$.

The general geometrical description [Marsden 1994] starts with the assumption
that an elastic material is a continuum $\mathbb{B}\subset\mathbb{R}^3$ which
may smoothly deform inside the space $\mathbb{R}^3$, so, it can be endowed with
differentiable structure, i.e. having an elastic material is formally
equivalent to have a smooth real 3-dimensional submanifold
$\mathbb{B}\subset\mathbb{R}^3$. The deformations are considered as smooth maps
(mostly embeddings) $\varphi: \mathbb{B}\rightarrow\mathbb{R}^3$. The spaces
$\mathbb{B}$ and $\mathbb{R}^3$ are endowed with riemannian metrics
$\mathbf{G}$ and $g$ respectively (and corresponding riemannian co-metrics
$\mathbf{G}^{-1}$ and $g^{-1}$), and induced isomorphisms
$\tilde{\mathbf{G}}$ and $\tilde{g}$
between the corresponding tangent and cotangent spaces .
This allows to define linear map inside every tangent space of
$\mathbb{B}$ in the following way: a tangent vector $V\in T_x\mathbb{B},\,
x\in\mathbb{B},$ is sent through the differential $d\varphi$ of $\varphi$ to
$(d\varphi)_x(V)\in T_{\varphi(x)}\mathbb{R}^3$, then by means of the
isomorphism $\tilde{g}$ we determine the corresponding 1-form (i.e. we "lower
the index"), this 1-form is sent to the dual space $T^*_x\mathbb{B}$ of
$T_x\mathbb{B}$ by means of the dual linear map $(d\varphi)^*:
T^*_{\varphi(x)}\mathbb{\mathbb{R}}^3\rightarrow T^*_x\mathbb{B}$, and finally,
we determine the corresponding tangent vector by means of the isomorphism
$\tilde{\mathbf{G}}^{-1}$ (i.e. we "raise the index" correspondingly). The so
obtained linear map
$$
\mathbf{C}_x :=
\big[\tilde{\mathbf{G}}^{-1}\circ
(d\varphi)^*\circ\tilde{g}\circ(d\varphi)\big]_x:
T_x\mathbb{B}\rightarrow T_x\mathbb{B}
$$
(which is denoted in [Marsden 1994] by $(\mathbf{F^TF})_x$), extended to the
whole $\mathbb{B}$, is called {\it Caushy-Green deformation tensor field}. Now,
the combination
$$
\mathbf{E}_x:=\frac12\big[(\tilde{\mathbf{G}}\circ\mathbf{C}-\mathbf{G})\big]_x=
\frac12\big[(d\varphi)^*\circ\tilde{g}\circ(d\varphi)-\mathbf{G}\big]_x:
T_x\mathbb{B}\times T_x\mathbb{B}\rightarrow \mathbb{R}
$$
is called {\it Lagrangian strain tensor field}. Following corresponding
linearization procedure (Ch.4 in [Marsden 1994]) defined by an appropriate
vector field $X$ on $\mathbb{B}$, representing an infinitesimal displacement of
$\mathbb{B}$, it can be shown that the linearization of $\mathbf{E}$ reduces to
$\frac12 L_X\,g$, where $L_X$ is the Lie derivative with respect to $X$.

We could look at the problem also as
follows. The mathematical counterparts of the allowed (including
reversible) deformations are the diffeomorphisms $\varphi$ of a riemannian
manifold $(M,g)$, and every $\varphi(M)$ represents a possible configuration of
the material considered. But some diffemorphisms do not lead to
deformation (i.e. to shape changes), so, a criterion must be introduced to
separate those diffeomorphisms which should be considered as essential. For
such a criterion is chosen the distance change: {\it if the distance between
any two fixed points does not change during the action of the external force
field, then we say that there is no deformation}. Now, every essential
diffeomorphism $\varphi$ must transform the metric $g$ to some new metric
$\varphi^*g$, such that $g\neq\varphi^*g$. The naturally arising tensor field
$e=(\varphi^*g-g)\neq 0$ appears as a measure of the physical abilities of the
material to withstand external force actions.

Since the external force is assumed to act locally and the material
considered gets the corresponding to the external force field final
configuration in a smooth way, i.e. passing smoothly through a family of
allowed configurations, we need a localization of the above scheme, such
that the isometry doffeomorphisms to be eliminated. This is done by means
of introducing 1-parameter group $\varphi_t, t\in [a,b]\subset\mathbb{R}$
of local diffeomorphisms, so, $\varphi_a(M)$ and $\varphi_b(M)$ denote
correspondingly the initial and final configurations. Now $\varphi_t$
generates a family of metrics $\varphi_t^*\,g$, and a corresponding family
of tensors $e_t$. According to the local analysis every local 1-parameter
group of diffeomorphisms is generated by a vector field on $M$. Let the
vector field $X$ generates $\varphi_t$. Then the quantity
$$
\frac12\,L_{X}g:=\frac12\,\lim_{t\rightarrow 0}\frac{\varphi_t^*\,g-g}{t} \ ,
$$
i.e. one half of the {\it Lie derivative} of $g$ along $X$, is called
(infinitesimal) {\it strain tensor}, or {\it deformation tensor}. \vskip
0.2cm {\bf Remark}. Further in the paper we shall work with $L_{X}\,g$,
i.e. the factor $1/2$ will be omitted. \vskip 0.2cm

In our further study we shall call $L_X\,g$, where $g=\eta$ is the
Minkowski (pseudo)metric, just {\it strain tensor}. We would like to note
that, as far as we know, photon-like objects have not been considered from
such a point of view. Clearly, the term "material" is not appropriate for
PhLO because no static situations are admissible,
{\bf our objects of interest are of entirely dynamical nature},
so the corresponding {\it relativistic strain} tensors must take care of this.

According to the previous section important vector fields in our approach
to describe electromagnetic PhLO are $\bar{\zeta},\,\bar{A},\,\bar{A^*}$,
so, we consider the corresponding three electromagnetic strain tensors: $
L_{\bar{\zeta}}\,\eta; \,L_{\bar{A}}\,\eta; \,L_{\bar{A^*}}\,\eta$. \vskip
0.3cm {\bf Proposition 4}. The following relations hold: \[
L_{\bar{\zeta}}\,\eta=0, \ \ \ (L_{\bar{A}}\,\eta)_{\mu\nu}\equiv
D_{\mu\nu}= \begin{Vmatrix}2u_x & u_y+p_x & u_z & u_{\xi} \\ u_y+p_x &
2p_y & p_z & p_{\xi} \\ u_z & p_z & 0 & 0 \\ u_{\xi} & p_{\xi} & 0 & 0 \ \
\ \end{Vmatrix} , \] \[ (L_{\bar{A^*}}\,\eta)_{\mu\nu}\equiv D^*_{\mu\nu}
= \begin{Vmatrix}-2\varepsilon p_x & -\varepsilon(p_y+u_x) & -\varepsilon
p_z & -\varepsilon p_{\xi} \\ -\varepsilon(p_y+u_x) & 2\varepsilon u_y &
\varepsilon u_z & \varepsilon u_{\xi} \\ -\varepsilon p_z & \varepsilon
u_z & 0 & 0 \\ -\varepsilon p_{\xi} & \varepsilon u_{\xi} & 0 & 0
\end{Vmatrix} . \]

{\bf Proof}. Immediately verified.
\vskip 0.3cm
We give now some important from our viewpoint relations.
\[
D(\bar{\zeta},\bar{\zeta})=D^*(\bar{\zeta},\bar{\zeta})=0,
\]
\[
D(\bar{\zeta})\equiv D(\bar{\zeta})_\mu dx^\mu\equiv D_{\mu\nu}\bar{\zeta}^\nu
dx^\mu =(u_\xi-\varepsilon u_z)dx + (p_\xi-\varepsilon p_z)dy,
\]
\[
D(\bar{\zeta})^\mu\frac{\partial}{\partial x^\mu}\equiv
D^{\mu}_{\nu}\bar{\zeta}^\nu\frac{\partial}{\partial x^\mu}=
-(u_\xi-\varepsilon u_z)\frac{\partial}{\partial x} -
(p_\xi-\varepsilon p_z)\frac{\partial}{\partial y}=-[\bar{A},\bar{\zeta}],\ \
\]
\[
D_{\mu\nu}\bar{A}^\mu\bar{\zeta}^\nu=
-\frac12\Big[(u^2+p^2)_\xi -\varepsilon(u^2+p^2)_z\Big]=
-\frac12L_{\bar{\zeta}}\phi^2 ,
\]
\[
D_{\mu\nu}\bar{A^*}^\mu\bar{\zeta}^\nu=
-\varepsilon\Big[u(p_\xi-\varepsilon p_z)-p(u_\xi-\varepsilon u_z)\Big]=
-\varepsilon\mathbf{R}=-\varepsilon \phi^2\,L_{\bar{\zeta}}\psi.
\]
We also have:
\[
D^*(\bar{\zeta})=\varepsilon\Big[-(p_\xi-\varepsilon p_z)dx+
(u_\xi-\varepsilon u_z)dy\Big] ,
\]
\[
D^*(\bar{\zeta})^\mu\frac{\partial}{\partial x^\mu}\equiv
(D^*)^{\mu}_{\nu}\bar{\zeta}^\nu\frac{\partial}{\partial x^\mu}=
-\varepsilon(p_\xi-\varepsilon p_z)\frac{\partial}{\partial x} +
(u_\xi-\varepsilon u_z)\frac{\partial}{\partial y}=[\bar{A^*},\bar{\zeta}],\ \
\]
\[
D^*_{\mu\nu}\bar{A^*}^\mu\bar{\zeta}^\nu=
-\frac12\Big[(u^2+p^2)_\xi -\varepsilon(u^2+p^2)_z\Big]=
-\frac12L_{\bar{\zeta}}\phi^2 ,
\]
\[
D^*_{\mu\nu}\bar{A}^\mu\bar{\zeta}^\nu=
\varepsilon\Big[u(p_\xi-\varepsilon p_z)-p(u_\xi-\varepsilon u_z)\Big]=
\varepsilon\mathbf{R}=\varepsilon \phi^2\,L_{\bar{\zeta}}\psi.
\]
Clearly, $D(\bar{\zeta})$ and $D^*(\bar{\zeta})$ are linearly independent in
general:
$$
D(\bar{\zeta})\wedge D^*(\bar{\zeta})=\varepsilon
\Big[(u_\xi-\varepsilon u_z)^2+(p_\xi-\varepsilon p_z)^2\Big]dx\wedge dy
=\varepsilon\phi^2(\psi_\xi-\varepsilon \psi_z)^2\,dx\wedge dy\neq 0.
$$
Recall now that every 2-form $F$ defines a linear map $\tilde{F}$ from
1-forms to 3-forms through the exterior product:
$\tilde{F}(\alpha):=\alpha\wedge F$, where $\alpha\in \Lambda^1(M)$.
Moreover, the Hodge $*$-operator, composed now with $\tilde{F}$, gets
$\tilde{F}(\alpha)$ back to $*\tilde{F}(\alpha)\in\Lambda^1(M)$. In the
previous section we introduced two 2-forms $G=A\wedge\zeta$ and
$G^*=A^*\wedge\zeta$ and noticed that $G^*=*G$. We readily obtain now
\[
D(\bar{\zeta})\wedge G=D^*(\bar{\zeta})\wedge G^*= D(\bar{\zeta})\wedge
A\wedge\zeta=D^*(\bar{\zeta})\wedge A^*\wedge\zeta=
\]
\[
=-\varepsilon\Big[u(p_\xi-\varepsilon p_z)- p(u_\xi-\varepsilon
u_z)\Big]dx\wedge dy\wedge dz- \Big[u(p_\xi-\varepsilon p_z)-
p(u_\xi-\varepsilon u_z)\Big]dx\wedge dy\wedge d\xi=
\]
\[
=-\phi^2\,L_{\bar{\zeta}}\psi\,(\varepsilon\,dx\wedge dy\wedge dz+
dx\wedge dy\wedge d\xi)=-\mathbf{R}\,(\varepsilon\,dx\wedge dy\wedge dz+
dx\wedge dy\wedge d\xi),
\]
\[ D(\bar{\zeta})\wedge
G^*=-D^*(\bar{\zeta})\wedge G= D(\bar{\zeta})\wedge
A^*\wedge\zeta=-D^*(\bar{\zeta})\wedge A\wedge\zeta=
\]
\[
=\frac12\Big[(u^2+p^2)_\xi-\varepsilon(u^2+p^2)_z\Big] (dx\wedge dy\wedge
dz+\varepsilon\,dx\wedge dy\wedge d\xi).
\]
Thus, recalling relations (12)-(15), we get
\begin{equation}
*\Big[D(\bar{\zeta})\wedge
A\wedge\zeta\Big]= *\Big[D^*(\bar{\zeta})\wedge A^*\wedge\zeta\Big]=
-\varepsilon\mathbf{R}\,\zeta =-i(\bar{G^*})\mathbf{d}G=i(\bar{G})\mathbf{d}G^*,
\end{equation}
\begin{equation} *\Big[D(\bar{\zeta})\wedge
A^*\wedge\zeta\Big]= -*\Big[D^*(\bar{\zeta})\wedge A\wedge\zeta\Big]=
\frac12L_{\bar{\zeta}}\phi^2\,\zeta=
i(\bar{G})\mathbf{d}G=i(\bar{G^*})\mathbf{d}G^*.
\end{equation}

The above relations show various dynamical aspects of the energy-momentum
redistribution during evolution of our PhLO. In particular, equations (16-17)
clearly show that it is possible the translational and rotational components of
the energy-momentum redistribution to be represented in form depending on the
$\zeta$-directed strains $D(\zeta)$ and $D^*(\zeta)$. So, the local
translational changes of the energy-momentum carried by the two vector
components $G$ and $G^*$ of our PhLO are given by the two 1-forms
$*\big[D(\bar{\zeta})\wedge A^*\wedge\zeta\big]$ and
$*\big[D^*(\bar{\zeta})\wedge A\wedge\zeta\big])$ and the local rotational ones
- by the 1-forms $*\big[D(\bar{\zeta})\wedge A\wedge\zeta\big]$ and
$*\big[D^*(\bar{\zeta})\wedge A^*\wedge\zeta\big]$.  In fact, the form
$*\big[D(\bar{\zeta})\wedge A\wedge\zeta\big]$ determines the strain that
"leaves" the 2-plane defined by $(A,\zeta)$ and the form
$*\big[D^*(\bar{\zeta})\wedge A^*\wedge\zeta\big]$ determines the strain that
"leaves" the 2-plane defined by $(A^*,\zeta)$. Since the PhLO is free, i.e. no
energy-momentum is lost or gained, this means that the two (null-field)
components $G$ and $G^*$ exchange locally {\it equal} energy-momentum
quantities:
 $i(\bar{G^*})\mathbf{d}G=-i(\bar{G})\mathbf{d}G^*$.
Moreover, the easily verified relation
$G_{\mu\sigma}G^{\nu\sigma}=(G^*)_{\mu\sigma}(G^*)^{\nu\sigma}$ shows that the
two components $G$ and $G^*$ carry the same stress-energy-momentum. Now, the
local energy-momentum conservation law
$\nabla_{\nu}\big[G_{\mu\sigma}\bar{G}^{\nu\sigma}+
(G^*)_{\mu\sigma}(\bar{G}^*)^{\nu\sigma}\big]=0$ requires
$L_{\bar{\zeta}}\phi^2=0$, and the corresponding strain-fluxes become
 zero: $*\big[D^*(\bar{\zeta})\wedge A\wedge\zeta\big]=0$,
$*\big[D(\bar{\zeta})\wedge A^*\wedge\zeta\big]=0$. On the other hand, only
dynamical relation between  the energy-momentum change and strain fluxes
exists, so NO analog of the assumed in elasticity theory generalized Hooke
law, (i.e. linear relation between the stress tensor and the strain tensor)
seems to exist. This clearly goes along with the fully dynamical nature of PhLO,
i.e. linear relations exist between the divergence terms of our stress tensor
$\frac12\big[-G_{\mu\sigma}\bar{G}^{\nu\sigma}-
(G^*)_{\mu\sigma}(\bar{G}^*)^{\nu\sigma}\big]$
and the $\bar{\zeta}$-directed strain fluxes as given by equations (16)-(17).

\section{The translational-rotational consistency}
We begin this section with summarizing from physical viewpoint some of the
results of the preceding two sections in the following \vskip 0.3cm {\bf
Corollary}. {\it An electromagnetic PhLO has two subsystems, mathematically
represented by the two 2-forms $G$ and $G^*$, these two subsystems carry the
same energy-momentum, and they are in a permanent dynamical equilibrium: each
one gives locally to the other as much energy-momentum as it gains locally from
it}. \vskip 0.3cm This conclusion and the considerations in the preceding two
sections allow to make explicit the mathematical representation of the PhLO
translational-rotational structure. In fact, it is seen that the rotational
component of propagation of our PhLO is dimensionally localized in a 2-plane,
which in our consideration is parametrized by coordinates $(x,y)$, and the
translational component of propagation is of constant nature and evolves along
$\bar{\zeta}$. Since our PhLO is of electromagnetic nature, the corresponding
energy-momentum tensor should be represented by
$\frac12\big[-G_{\mu\sigma}\bar{G}^{\nu\sigma}-
(G^*)_{\mu\sigma}(\bar{G}^*)^{\nu\sigma}\big]$.
Now, the corresponding local energy-momentum conservation law
$\nabla_{\nu}\big[G_{\mu\sigma}\bar{G}^{\nu\sigma}+
(G^*)_{\mu\sigma}(\bar{G}^*)^{\nu\sigma}\big]=0$
reduces to the dynamical equation
$L_{\bar{\zeta}}\phi^2=L_{\bar{\zeta}}(u^2+p^2)=0$,
which seems to be naturally accepted to represent the translational
component of propagation.

In order to come to some appropriate dynamical picture of the rotational
component of propagation we make the following consideration. Recall the two
vector fields $\bar{A}$ and $\bar{A^*}$. Since
$\bar{A}\wedge\bar{A^*}=-
\varepsilon(u^2+p^2)\,\partial_x\wedge \partial_y\neq 0$, then at
all space-time points occupied by our PhLO we have the frame
$\Sigma_1=(\bar{A}, \bar{A^*},
\partial_z, \partial_\xi)$. On the other hand the vector
fields $[\bar{A},\bar{\zeta}]$ and $[\bar{A^*},\bar{\zeta}]$ are also linearly
independent in general:
$$
[\bar{A},\bar{\zeta}]\wedge[\bar{A^*},\bar{\zeta}]=
\varepsilon\big[(u_\xi-\varepsilon u_z)^2+(p_\xi-\varepsilon p_z)^2\big]
\frac{\partial}{\partial x}\wedge\frac{\partial}{\partial y}
=\varepsilon\phi^2(\psi_\xi-\varepsilon\psi_z)^2
\frac{\partial}{\partial x}\wedge\frac{\partial}{\partial y} \, ,
$$
and we get the new frame
$\Sigma_2=([\bar{A},\bar{\zeta}],[\bar{A^*},\bar{\zeta}],\partial_z,
\partial_\xi)$, which is regular only if $\mathbf{R}\neq 0$.
Hence, the rotational component of propagation transforms
the frame $\Sigma_1$ to the frame $\Sigma_2$. We see that essentially, the
2-frame $(\bar{A},\bar{A^*})$ is transformed to the 2-frame
$([\bar{A},\bar{\zeta}], [\bar{A^*},\bar{\zeta}])$, and these two 2-frames are
tangent to the $(x,y)$-plane. So we get the linear map

\[
\big([\bar{A},\bar{\zeta}], [\bar{A^*},\bar{\zeta}]\big)=
\big(\bar{A}, \bar{A^*}\big)
\begin{Vmatrix}\alpha & \beta \\ \gamma & \delta\end{Vmatrix} .
\]
Solving this system with respect to $(\alpha,\beta,\gamma,\delta)$ we obtain
\[
\begin{Vmatrix}\alpha & \beta \\ \gamma & \delta \end{Vmatrix}=
\frac{1}{\phi^2}
\begin{Vmatrix} -\frac12 L_{\bar{\zeta}}\phi^2 &
\varepsilon \mathbf{R} \\ -\varepsilon \mathbf{R}
& -\frac12 L_{\bar{\zeta}}\phi^2 \end{Vmatrix}=
-\frac12\frac{L_{\bar{\zeta}}\phi^2}{\phi^2}
\begin{Vmatrix} 1 & 0 \\ 0 & 1\end{Vmatrix}+
\varepsilon L_{\bar{\zeta}}\psi\begin{Vmatrix} 0 & 1 \\ -1 & 0\end{Vmatrix} .
\]
Assuming the conservation law $L_{\bar{\zeta}}\phi^2=0$, we obtain that
the rotational component of propagation is governed by the matrix
$\varepsilon L_{\bar{\zeta}}\psi\,J$, where $J$ denotes the canonical
complex structure in $\mathbb{R}^2$, and since
$\phi^2\,L_{\bar{\zeta}}\psi=\mathbf{R}$ we conclude that the rotational
component of propagation is available if and only if the Frobenius
curvature is NOT zero: $\mathbf{R}\neq 0$. We may also say that a
consistent translational-rotational dynamical structure is available if
the amplitude $\phi^2=u^2+p^2$ is a running wave along $\bar{\zeta}$ and
the phase $\psi=\mathrm{arctg}\frac{p}{u}$ is NOT a running wave along
$\bar{\zeta}$.

As we noted before the local conservation law $L_{\bar{\zeta}}\phi^2=0$,
being equivalent to $L_{\bar{\zeta}}\phi=0$, gives one dynamical linear
first order equation. This equation pays due respect to the assumption
that our spatially finite PhLO, together with its energy density,
propagates translationally with the constant velocity $c$. We need one
more equation in order to specify the phase function $\psi$. If we pay
corresponding respect also to the rotational aspect of the PhLO nature it
is desirable this equation {\it to introduce and guarantee the
conservative and constant character of this aspect of PhLO nature}. Since
rotation is available only if $L_{\bar{\zeta}}\psi\neq 0$, the simplest such
assumption respecting the constant character of the rotational component of
propagation seems to be $L_{\bar{\zeta}}\psi=const\neq 0$. Now, since the usual
physical dimension of the canonical coordinates $(x,y,z,\xi)$ is [length] and
the phase $\psi$ is dimensionless, we may put
$L_{\bar{\zeta}}\psi=const=\kappa/l_o$, where $\kappa=\pm 1$ and $l_o$ is a
positive constant with [$l_o$]=[length]. Note that $l_o$ is equal to the
square root of the relation of the volumes defined by the frames $\Sigma_2$ and
$\Sigma_1$ : $l_o=\sqrt{|vol(\Sigma_2)/vol(\Sigma_1)|}=
\sqrt{|\omega_o(\Sigma_2)/\omega_o(\Sigma_1)|}$.

Thus, the equation
$L_{\bar{\zeta}}\phi=0$ and the frame rotation
$[\bar{A},\bar{\zeta}]=-\varepsilon\bar{A^*}\,L_{\bar{\zeta}}\psi$
and $[\bar{A^*},\bar{\zeta}]=\varepsilon\bar{A}\,L_{\bar{\zeta}}\psi $,
i.e.
$(\bar{A},\bar{A^*},\partial_z,
\partial_\xi)\rightarrow
([\bar{A},\bar{\zeta}],[\bar{A^*},\bar{\zeta}],\partial_z, \partial_\xi)$,
give the following equations for the two functions $(u,p)$:

\[
u_\xi-\varepsilon u_z=-\frac{\kappa}{l_o}\,p, \ \ \
p_\xi-\varepsilon p_z=\frac{\kappa}{l_o}\,u \ .
\]
If we now introduce the complex valued function
$\Psi=u\,I+p\,J$, where $I$ is the identity map in $\mathbb{R}^2$,
the above two equations are equivalent to
$$
L_{\bar{\zeta}}\Psi=\frac{\kappa}{l_o}J(\Psi) \ ,
$$
which clearly confirms once again the translational-rotational consistency
in the form that {\it no translation is possible without rotation, and no
rotation is possible without translation}, where the rotation is
represented by the complex structure $J$. Since the operator $J$ rotates
to angle $\alpha=\pi/2$, the parameter $l_o$ determines the corresponding
translational advancement, and $\kappa=\pm 1$ takes care of the left/right
orientation of the rotation. Clearly, a full rotation (i.e.
$2\pi$-rotation) will require a $4l_o$-translational advancement, so, the
natural time-period is $T=4l_o/c=1/\nu$, and $4l_o$ is naturally interpreted as
the PhLO size along the spatial direction of translational propagation.

In order to find an integral characteristic of the PhLO rotational nature
in {\it action units} we correspondingly modify, (i.e. multiply by
$l_o/c$) and consider any of the two equal Frobenius curvature
generating 4-forms:
$$
\frac{l_o}{c}\,\mathbf{d}A\wedge A\wedge \zeta=
\frac{l_o}{c}\,\mathbf{d}A^*\wedge A^*\wedge \zeta=
\frac{l_o}{c}\,\varepsilon \mathbf{R}\omega_o=
\frac{l_o}{c}\,\varepsilon\phi^2\,L_{\bar{\zeta}}\psi\,\omega_o=
\varepsilon\kappa\frac{l_o}{c}\frac{\phi^2}{l_o}\,\omega_o=
\varepsilon\kappa\frac{\phi^2}{c}\,\omega_o .
$$
Integrating this 4-form over the 4-volume $\mathbb{R}^3\times [0,4l_o]$ we
obtain the quantity $\mathcal{H}=\varepsilon\kappa ET=\pm\,ET$, where $E$ is
the integral energy of the PhLO and $T=4l_o/c$, which clearly is the analog of
the Planck formula $E=h\nu$, i.e. $h=ET$.

\section{Lagrangian formulation: a complex valued scalar field}

 Consider the space
$\mathbb{R}^4=\mathbb{R}^3\times\mathbb{R}$, where $\mathbb{R}$ determines
the time-direction. Let the field of complex numbers
$\mathbb{C}=(\mathbb{R}^2,J),\, J\circ J=-id_{\mathbb{R}^2}$ be given a real
representation as a 2-dimensional real vector space with basis
\[
I=\begin{Vmatrix}1 & 0 \\0 & 1\end{Vmatrix}, \ \ J=\begin{Vmatrix}0 & 1 \\-1
& 0\end{Vmatrix}.
\]
Every $\mathbb{C}$-valued function $\alpha$ on $\mathbb{R}^4$ can be
represented in the form $\alpha=uI+pJ=\phi\cos\psi\,I+\phi\sin\psi\,J$,
where $u$ and $p$ are two real-valued functions, $\phi=\sqrt{u^2+p^2}$ and
$\psi=\arctan\frac pu$, and the components of $\alpha$ with respect to
this basis will be numbered by latin indices taking values $(1,2):
\alpha_i, i=1,2$ . We denote further $J(\alpha)=-pI+uJ\equiv \bar{\alpha}$
and we shall make use of the introduced in the preceding section constant
$l_o>0$ . Now we introduce two new 1-forms: the first one, denoted by
$k^s$, is the restriction of $\frac1l_o\zeta$ to $\mathbb{R}^3$ and then
extended to the whole $\mathbb{R}^4$ through zero-components in the
$\bar{\zeta}$-adapted coordinate system: $k^s=\frac{\varepsilon}{l_o}dz$;
the second one, denoted by $k^\xi$, is the restriction of $\frac1l_o\zeta$
to the time-direction $\mathbb{R}$ and then extended to the whole
$\mathbb{R}^4$ in the same way:  $k^\xi=\frac1l_o d\xi$. Hence, in the
$\bar{\zeta}$-adapted coordinate system we obtain
$k^s=(0,0,\varepsilon/l_0,0)$, and $k^\xi=(0,0,0,1/l_o)$.

Making use of our vector field $\bar{\zeta}$, of the inner product
$g$ in $\mathbb{C}=(\mathbb{R}^2,J)$ defined by
$g(\alpha,\beta)=\frac12tr(\alpha\circ\beta^*)$, $\beta^*$ is the transposed
to $\beta$, we consider now the following lagrangian (summation over the
repeated indices: $g(\alpha,\alpha)=\alpha_i\alpha_i$):
\[
\mathbb{L}=
\frac12\Big[\kappa l_o g(\alpha,L_{\bar{\zeta}}\bar{\alpha})+g(\alpha,\alpha)-
\kappa l_o
g(\bar{\alpha},L_{\bar{\zeta}}\alpha)+g(\bar{\alpha},\bar{\alpha})\Big]=
\]
\[
=\frac12\Big[\alpha_i\Big(\kappa l_o\bar{\zeta}^\sigma\frac{\partial
\bar{\alpha}_i}{\partial x^\sigma}+\alpha_i\Big)-
\bar{\alpha}_i\Big(\kappa
l_o\bar{\zeta}^\sigma\frac{\partial\alpha_i}{\partial
x^\sigma}-\bar{\alpha}_i\Big)\Big],
\]
where $\kappa=\pm 1$. Considering $\alpha$ and $\bar{\alpha}$ as independent,
the Lagrange equations read
\[
\kappa l_o\bar{\zeta}^\sigma\frac{\partial \bar{\alpha}_i}{\partial
x^\sigma}=-\alpha_i \ ;\ \
\kappa l_o\bar{\zeta}^\sigma\frac{\partial \alpha_i}{\partial
x^\sigma}=\bar{\alpha}_i.
\]
Note that the first equation follows from the second one under the action
with $J$ from the left, hence, we have just one equation of the form
$\kappa l_o L_{\bar{\zeta}}\alpha=J(\alpha)$, which represents the idea
for {\it consistent translational-rotational propagation}: the
translational change $L_{\bar{\zeta}}\alpha$ of the field $\alpha$ is
proportional to the rotational change $J(\alpha)$ and the coefficient
$l_o$ gives the translational advancement for a rotation of $\pi/2$.

In terms of
$\phi=\sqrt{u^2+p^2}$ and  $\psi=\mathrm{arctan}\frac{p}{u}$ these
equations give
\[
L_{\bar{\zeta}}\phi.\cos(\psi)
-\phi.\sin(\psi)\left(L_{\bar\zeta}\psi-
\frac{\kappa}{l_o}\right)=0, \ \
L_{\bar{\zeta}}\phi.\sin(\psi)
+\phi.\cos(\psi)\left(L_{\bar\zeta}\psi-
\frac{\kappa}{l_o}\right)=0.
\]
These two equations are consistent only if
\begin{equation}
L_{\bar{\zeta}}\phi=0,\ \
L_{\bar{\zeta}}\psi=\frac{\kappa}{l_o}.         
\end{equation}
The solutions are:
\begin{equation}
\phi=\phi(x,y,\xi+\varepsilon z); \ \
\psi_1=-\varepsilon
\frac{\kappa}{l_o} z +f(x,y,\xi+\varepsilon z); \     
\ \psi_2=
\frac{\kappa}{l_o}\xi+f(x,y,\xi+\varepsilon z),
\end{equation}
where $f$ is an arbitrary function. Assuming $f=const$ we see that
$$
\psi_1=-\varepsilon\frac{\kappa}{l_o} z+const=-\kappa\,k^s_\mu x^\mu+const
\ \ \ \text{and}
\ \ \ \psi_2=\frac{\kappa}{l_o} \xi+const =\kappa\,k^\xi_\mu x^\mu+const
$$
are the simplest possible solutions leading to non-zero curvature. We note
that the spatial structure of the solution defined by $\psi_1$ is {\it
phase dependent} while the spatial structure of the solution defined by
$\psi_2$ is NOT  phase dependent.

We note that this lagrangian leads to the obtained in the previous section {\it
linear} equations for the components of $\alpha$, which equations admit
3d-finite solutions of the kind $$ \alpha_1=\phi\cos\psi ; \ \
\alpha_2=\phi\sin\psi $$ with consistent translational-rotational behavior,
where $\phi$ and $\psi$ are given above, and $\phi$ is a spatially finite
function.

It is easily seen that the lagrangian becomes zero on the solutions, and
since this lagrangian does NOT depend on any space-time metric the
corresponding Hilbert energy-momentum tensor is zero on the solutions.
This special feature of the lagrangian requires to look for another
procedure leading to corresponding conserved quantities. A good candidate
seems to be $T^{\mu\nu}=\phi^2 \bar{\zeta}^\mu\bar{\zeta}^\nu$.
In fact, we obtain (in our coordinate system) \[ \nabla_\nu
T^{\mu\nu}=\bar{\zeta}^\mu\nabla_\nu(\phi^2\bar{\zeta}^\nu)+ \phi^2
\bar{\zeta}^\nu\nabla_{\nu}\bar{\zeta}^\mu= \bar{\zeta}^\mu
L_{\bar{\zeta}}(\phi^2)+ \phi^2
\bar{\zeta}^\nu\nabla_{\nu}\bar{\zeta}^\mu. \] The first term on the right
is equal to zero on the solutions and the second term is zero since the vector
field $\bar{\zeta}$ is autoparallel, so, $\nabla_\nu T^{\mu\nu}=0$.

\section{Lagrangian formulation: exterior 2-form field}
We show now that the same equations for the two functions $(u,p)$, or
$(\phi,\psi)$, can be obtained from a lagrangian defined in terms of a
2-form. Recall that the space $\Lambda^2(\mathbb{R}^4)$ of 2-forms on
$\mathbb{R}^4$ is 6-dimensional and denote by $\mathcal{J}$ the complex
structure in this space defined by:
$\mathcal{J}_{16}=-\mathcal{J}_{25}=\mathcal{J}_{34}=-
\mathcal{J}_{43}=\mathcal{J}_{52}=-\mathcal{J}_{61}=1$, and all other
components of  $\mathcal{J}$ are zero in our $\bar{\zeta}$-adapted
coordinate system, i.e. the only non-zero elements are the off-diagonal
components, and they alternatively change from $(+1)$ (upper right angle)
to $(-1)$ (lower left angle).

We define now a representation $\rho$ of the algebra
$\mathbb{C}$ in the algebra $L_{\Lambda^2(\mathbb{R}^4)}$ of linear
maps in the 2-forms on $\mathbb{R}^4$ by the relation
\begin{equation}
\rho(\alpha_\varepsilon)=\rho(uI+\varepsilon pJ)\overset{\text{\small def}}{=}
u\mathcal{I}+\varepsilon p\mathcal{J}, \ \                     
\mathcal{I}=id_{\Lambda^2(\mathbb{R}^4)},
\ \ \alpha_{\varepsilon}\in\mathbb{C}, \ \ \varepsilon=\pm 1.
\end{equation}
Clearly,
$\rho(\alpha+\beta)=\rho(\alpha)+\rho(\beta), \  \rho(\alpha.\beta)=
\rho(\alpha)\circ\rho(\beta)$, and if $G$ is an arbitrary 2-form then
$\rho(\alpha_\varepsilon).G=uG+\varepsilon p\mathcal{J}(G)$.
Note that here and further in the text the
couple $(u,p)$ may denote the complex number $(uI+pJ)$, as well as the
complex-valued function $\alpha=u(x,y,z,\xi)I+p(x,y,z,\xi)J$.

Let's go back now to our $\bar\zeta$-adapted coordinate system and
consider the 2-form $F_o=dx\wedge\zeta=dx\wedge (\varepsilon dz+d\xi)=
\varepsilon dx\wedge dz+dx\wedge d\xi$. Recalling the two 1-forms
$A=udx+pdy$ and $A^*=-pdx+udy$ (we omit $\varepsilon$ before $p$ and $u$
as it was defined in Sec.2) we obtain \[
\rho(\alpha_{\varepsilon}).F_o=\varepsilon udx\wedge dz+ \varepsilon
pdy\wedge dz+udx\wedge d\xi+pdy\wedge d\xi=A\wedge\zeta, \] \[
\rho(J(\alpha_{\varepsilon})).F_o= (-\varepsilon
p\mathcal{I}+u\mathcal{J}).F_o= \] \[ =-pdx\wedge dz+ udy\wedge
dz-\varepsilon pdx\wedge d\xi+\varepsilon udy\wedge d\xi =A^*\wedge\zeta=
\mathcal{J}(\rho(\alpha_\varepsilon).F_o). \] Since $\rho$ is a linear map
and $\rho(0)=0$, we get one-to-one map between the $\mathbb{C}$-valued
functions on $\mathbb{R}^4$ and a special subset of 2-forms. All such
2-forms depend on the choice of the 1-form $\zeta$, while the dependence
on $dx$ is not essential.  Also, they are isotropic: $$
(A\wedge\zeta)\wedge (A\wedge\zeta)=0, \ (A\wedge\zeta)\wedge
\mathcal{J}(A\wedge\zeta)= (A\wedge\zeta)\wedge (A^*\wedge\zeta)=0, $$
i.e. they have zero invariants.

Moreover, every such 2-form may be considered as a linear map in
$\Lambda^2(\mathbb{R}^4)$ through the above correspondence:
$\rho(\alpha_\varepsilon).F_o\rightarrow \rho(\alpha_\varepsilon)$. Since
together with the zero element of $\Lambda^2(M)$ these 2-forms define a
linear space $V_\zeta$, this property suggests to introduce inner product
in this linear space by the rule $$ <G^1_\varepsilon (a,b),
G^2_\varepsilon (m,n)>= \frac16
tr\Big[\rho\big[\alpha_\varepsilon(m,n)\big]\circ
\rho\big[\alpha^*_\varepsilon (a,b)\big]\Big]=am+bn. $$ Hence, every such
2-form acquires a norm.

Let now $F$ and $G$ be two arbitrary 2-forms. In order to define the
lagrangian we consider the Minkowski space-time $M=(\mathbb{R}^4,\eta)$ as a
real manifold, where the pseudoeuclidean metric $\eta$ has signature
$(-,-,-,+)$, and make use of the Lie derivative $L_{\bar\zeta}$ with respect
to the vector field $\bar\zeta$. Also, $-(k^s)^2=(k^\xi)^2=(l_o)^{-2}$.
Consider the lagrangian
\[
\mathbb{L}=\eta\left(\kappa l_oL_{\bar{\zeta}}G+F,F\right)-
\eta\left(\kappa l_oL_{\bar{\zeta}}F-G,G\right).
\]
In components in the $\zeta$-adapted coordinates where $\zeta^\sigma=const$, we
can write
\begin{equation} \mathbb{L}=\frac12\left(\kappa l_o\bar{\zeta}^\sigma
\frac{\partial G_{\alpha\beta}}{\partial x^\sigma}+
F_{\alpha\beta}\right)F^{\alpha\beta}-
\frac12\left(\kappa l_o\bar{\zeta}^\sigma                          
\frac{\partial F_{\alpha\beta}}{\partial x^\sigma}-
G_{\alpha\beta}\right)G^{\alpha\beta}, \ \ 0<l_o=const, \ \ \kappa=\pm 1,
\end{equation}
where $F^{\alpha\beta}=\eta^{\alpha\mu}\eta^{\beta\nu}F_{\mu\nu}$ and
$G^{\alpha\beta}=\eta^{\alpha\mu}\eta^{\beta\nu}G_{\mu\nu}$. Note that this
lagrangian is invariant with respect to $(F,G)\rightarrow(G,-F)$, or to
$(F,G)\rightarrow(-G,F)$.
The corresponding equations read
\[
\kappa l_o\bar{\zeta}^\sigma
\frac{\partial G_{\alpha\beta}}{\partial x^\sigma}+
F_{\alpha\beta}=0, \ \
\kappa l_o\bar{\zeta}^\sigma
\frac{\partial F_{\alpha\beta}}{\partial x^\sigma}-
G_{\alpha\beta}=0.
\]
In coordinate-free form these equations look like
\[
\kappa l_oL_{\bar{\zeta}}G=-F, \ \ \ \kappa l_oL_{\bar{\zeta}}F=G.
\]

Recall now the complex structure $\mathcal{J}$ and assume $F=-\mathcal{J}(G)$.
Then, treating $G$ and $\mathcal{J}(G)$ as independent (in fact they are
linearly independent on the real manifold $M$), and in view of of the constancy
of $\mathcal{J}$ in the $\zeta$-adapted coordinates, we can write
\begin{equation}
\kappa l_oL_{\bar{\zeta}}\mathcal{J}(G)=-G, \ \ \ \kappa
l_oL_{\bar{\zeta}}G=\mathcal{J}(G) .                          
\end{equation}
Since in our coordinates $\mathcal{J}$ and $\bar\zeta$ have constant
coefficients, clearly, $L_{\bar\zeta}$ and $\mathcal{J}$ commute, so that
the second (first) equation is obtained by acting with $\mathcal{J}$ from
the left on the first (second) equation, i.e. the above mentioned
invariance with respect to the transformations $(F,G)\rightarrow(\pm G,\mp
F)$ is reduced to $\mathcal{J}$-invariance.
Restricting now to 2-forms of the above defined kind and
recalling the way how
$\mathcal{J}$ acts, $\mathcal{J}(A\wedge\zeta)=A^*\wedge\zeta$, i.e. the
couple $(A,\zeta)$ is rotated to the couple $(A^*,\zeta)$, we naturally
interpret the last equations (22) as realization of the {\it
translational-rotational consistency}: the translational change of $G$
along $\bar\zeta$ is proportional to the rotational change of $G$
determined by $\mathcal{J}$, so, roughly speaking, {\bf no
$\bar\zeta$-translation ($\mathcal{J}$-rotation) is possible without
$\mathcal{J}$-rotation ($\bar\zeta$-translation)}, and the
$\mathcal{J}$-rotation corresponds to $l_o$ translational advancement.

From (22) it follows that on the solutions the lagrangian becomes zero.
So, if we try to define the corresponding Hilbert energy-momentum tensor
the variation of the volume element with respect to $\eta$ is not
essential. Moreover, the special quadratic dependence of $\mathbb{L}$ on
$\eta$ shows that the variation of $\mathbb{L}$ with respect to $\eta$
will also become zero on the solutions. Hence, this is another example of
the non-universality of the Hilbert method to define appropriate
energy-momentum tensor. As for the canonical energy-momentum tensor, it is
not symmetric, and its symmetrization is, in some extent, an arbitrary
act, therefore, we shall not make use of it.

We continue to restrict the equations (22) onto the subset of 2-forms $G$ of
the kind $G=\rho(\alpha_\varepsilon).F_o$. As it was mentioned all these
2-forms have zero invariants:
$G_{\mu\nu}G^{\mu\nu}=G_{\mu\nu}(\mathcal{J}(G))^{\mu\nu}=0$, or in
coordinate-free way, $G\wedge G=G\wedge\mathcal{J}(G)=0$. Moreover, the
easily verified relations $i(\bar\zeta)G=i(\bar\zeta)\mathcal{J}(G)=0$
show an {\it intrinsic} connection to $\bar\zeta$: it is the only
isotropic eigen vector of $G_\mu^\nu=\eta^{\mu\sigma}G_{\nu\sigma}$ and
$(\mathcal{J}G)_\mu^\nu=\eta^{\mu\sigma}(\mathcal{J}G)_{\nu\sigma}$.

Substituting $G=\rho(\alpha_\varepsilon).F_o$ we get the
already known equations
\begin{equation}
\kappa l_o (u_\xi-\varepsilon u_z)=-p, \ \ \          
\kappa l_o(p_\xi-\varepsilon p_z)=u \ .
\end{equation}
Clearly, $\phi^2\zeta\otimes\bar\zeta$ is the right choice for energy-momentum
tensor.

Note that the 2-form $F_o=dx\wedge\zeta$ satisfies the
 equation $L_{\bar{\zeta}}\phi=0$ since $\phi_{F_{o}}=1$,
and does NOT satisfy the equation for $\psi$, since
$\psi_{F_o}=0,2\pi,4\pi, ...$, so, $L_{\bar\zeta}(\psi_{F_o})=0$. In view
of this further we consider only not-constant $\mathbb{C}$-valued
functions.

As we already mentioned an appropriate local representative of the
rotational properties of these solutions appears to be any of the two
Frobenius 4-forms $\mathbf{d}A\wedge A\wedge\zeta $ and
$\mathbf{d}A^*\wedge A^*\wedge\zeta $, multiplied by the coefficient
$l_o/c$, so that integrating over the 4-region $(\mathbb{R}^3\times 4l_o)$
we get $\pm ET$, which carries integral information about the rotational
properties of the solution.

We'd like to mention also that the 3-forms
$
i(\bar{\zeta})(\mathbf{d}A\wedge A\wedge\zeta)=
i(\bar{\zeta})(\mathbf{d}A^*\wedge A^*\wedge\zeta),
$
which in our coordinate system look like $\gamma\wedge\zeta$ with
$\gamma=-\phi^2(L_{\bar{\zeta}}\,\psi)\, dx\wedge dy$, are closed.

The linear character of the equations obtained sets the question if the
superposition principle holds. In general, let the parameters
$\kappa, \varepsilon, l_o$ of the two solutions be different. Let now
$F_1(\kappa_1,\varepsilon_1,l_o^1;u,p)$ and
$F_2(\kappa_2,\varepsilon_2,l_o^2;m,n)$ be two solutions along the same
direction defined by $\bar\zeta$, and $\varepsilon$ of $\zeta$ is of
course equal to $\varepsilon_1$ for the first solution, and equal to
$\varepsilon_2$ for the second solution.  We ask now whether the linear
combination $c_1F_1+c_2F_2$ with $c_1=const, c_2=const$ will be also a
solution
$F_3(\kappa_3,\varepsilon_3,l_o^3;c_1u+c_2m,c_1p+c_2n)$
along the same direction?  In order this to happen the following
equations must be consistent:
\[
\kappa_1\varepsilon_1l_o^1L_{\bar\zeta}u=-\varepsilon_1 p, \ \
\kappa_1\varepsilon_1l_o^1L_{\bar\zeta}p=\varepsilon_1 u, \ \
\kappa_2l_o^2L_{\bar\zeta}m=-\varepsilon_2 n, \ \
\kappa_2l_o^2L_{\bar\zeta}n=\varepsilon_2 m
\]
\[
\kappa_3\varepsilon_3l_o^3L_{\bar\zeta}(c_1u+c_2m)=-\varepsilon_3(c_1p+c_2n),
\ \ \kappa_3\varepsilon_3l_o^3L_{\bar\zeta}(c_1p+c_2n)=
\varepsilon_3 (c_1u+c_2m),
\]
where $\varepsilon_3$ is equal $\varepsilon_1$, or to $\varepsilon_2$.
The
corresponding consistency condition looks as follows:
\[
\kappa_3\varepsilon_3 l_o^3
=\frac{c_1p+c_2n} {\frac{\varepsilon_1\kappa_1c_1}{l_o^1}p+
\frac{\varepsilon_2\kappa_2c_2}{l_o^2}n}=
\frac{c_1u+c_2m}
{\frac{\varepsilon_1\kappa_1c_1}{l_o^1}u+
\frac{\varepsilon_2\kappa_2c_2}{l_o^2}m}.
\]
For example, the relations $\varepsilon_3\kappa_3l_o^3=
\varepsilon_2\kappa_2l_o^2=\varepsilon_1\kappa_1l_o^1$ are sufficient for
this superposition to be a solution. This means that if the two solutions
propagate translationally for example from $-\infty$ to $+\infty$, i.e.
$\varepsilon_1=\varepsilon_2=-1$, if the rotational orientations coincide,
i.e. $\kappa_1=\kappa_2$, and if the spatial periodicity parameters are
equal: $l_o^1=l_o^2=l_o^3$, the sum $(c_1F_1+c_2F_2)$ gives a solution.
In general, however, the combination $c_1F_1+c_2F_2$ will not be a
solution.

On the other hand, we can introduce a multiplicative structure in the
solutions of the kind $\rho(\alpha_\varepsilon).F_o$. In fact, if
$F_1(\kappa_1)=\rho(\alpha^1(\varepsilon_1)).F_o$ and
$F_2(\kappa_2)=\rho(\alpha^2(\varepsilon_2)).F_o$ are two such solutions
we define their product $F=F_1.F_2$ by $
F=F_1.F_2=\rho(\alpha^1.\alpha^2).F_o$.
Clearly, the amplitude of $F$ is
a product of the amplitudes of $F_1$ and $F_2$:
$\phi_F=\phi_{F_1}.\phi_{F_2}$ and the phase of $F$ is the sum of the
phases of $F_1$ and $F_2$: $\psi_F=\psi_{F_1}+\psi_{F_2}$. Now, $F$ will
be a solution only if $$ \varepsilon_1=\varepsilon_2=\varepsilon_{F}, \ \
\frac{\kappa_{F}}{l_o^{F}}=\frac{\kappa_1l_o^2+\kappa_2l_o^1}{l_o^1.l_o^2}. $$
From the first of these relations it follows that in order $F$ to be a
solution, $F_1$ and $F_2$ must NOT move against each other, and then the
product-solution shall follow the same translational direction. However, it
is allowed $F_1$ and $F_2$ to have different rotational orientations, i.e.
$\kappa_1\neq\kappa_2$, then the product-solution will have rotational
orientation $\kappa_F=sign(\kappa_1l_o^2+\kappa_2l_o^1)$. Clearly, every
subset of solutions with the same $(\varepsilon,\kappa,l_o)$
 form a group with neutral element $F_o=\rho(I).F_o$, and
$F_\alpha^{-1}=\rho(\alpha^{-1}).F_o$,  moreover, the multiplicative group of
$\mathbb{C}$ acts on these solutions:  $(\alpha_\varepsilon,
F_\varepsilon)\rightarrow \rho(\alpha_\varepsilon).F_\varepsilon$.

We give now an explicit solution, i.e. a procedure to  construct a solution.
We assume that the phase is given by $\psi_1$
and $\varepsilon$ and $\kappa$ take values $\pm1$ independently. The form
of this solution $u=\phi\,\mathrm{cos}\,\psi_1;
\ \ p=\phi\,\mathrm{sin}\,\psi_1$
shows that the initial condition is determined entirely
by the choice of $\phi$, and it suggests also to choose the initial
condition $\phi_{t=0}(x,y,\varepsilon z)$ in the following way. Let for
$z=0$ the initial condition $\phi_{t=0}(x,y,0)$ be located on a disk
$D=D(x,y;a,b;r_o)$ of small radius $r_o$, the center of the disk to have
coordinates $(a,b)$, and the value of $\phi_{t=0}(x,y,0)$ to be
proportional to the distance $R(x,y,0)$ between the origin of the
coordinate system and the point $(x,y,0)$, so, $R(x,y,0)=\sqrt{x^2+y^2}$,
and $D$ is defined by $D=\{(x,y)|\sqrt{(x-a)^2+(y-b)^2}\leq r_o\}$. Also,
let $\theta_D$ be the smoothed out characteristic function of the disk
$D$, i.e. $\theta_D=1$ everywhere on $D$ except a very thin hoop-like zone
$B_D\subset D$ close to the boundary of $D$ where $\theta_D$ rapidly goes
from 1 to zero (in a smooth way), and $\theta_D=0$ outside $D$. Let also
the dependence on $z$ be given by
be the corresponding characteristic function $\theta(z;4l_o)$ of an
interval $(z,z+4l_o)$ of length $4l_o$ on the $z$-axis. If
$\gamma$ is the proportionality coefficient we obtain
$$
\phi(x,y,z,ct+\varepsilon z)=
\gamma.R(x,y,0).\theta_D.\theta(ct+\varepsilon z;4l_o).
$$
We see that because of the available {\it sine} and {\it cosine} factors
in the solution, the initial condition for the solution will occupy a
helical cylinder of height $4l_o$, having internal radius of $r_o$ and
wrapped up around the $z$-axis. Also, its center will always be
$R(a,b,0)$-distant from the $z$-axis. Hence, the solution will propagate
translationally along the coordinate $z$ with the velocity $c$, and
rotationally inside the corresponding infinitely long helical cylinder
because of the $z$-dependence of the available periodical multiples. The
curvature $K$ and the torsion $T$ of the screwline through the point
$(x,y,0)\in D$ will be \[ K=\frac{R(x,y,0)}{R^2(x,y,0)+b^2},\ \ \ \
T=\frac{\kappa\,b}{R^2(x,y,0)+b^2} \ , \] where $b=2l_o/\pi$. The
rotational frequency $\nu$ will be $\nu=c/2\pi b$, so we can introduce
period $T=1/\nu$ and elementary action $h=E.T$, where $E$ is the
(obviously finite) integral energy of the solution defined as 3d-integral
of the energy density $\phi^2$.

\section{Discussion and Conclusion}
The two basic features of our approach are the assumptions for continuous
spatially finite structure, and for available consistent
translational-rotational dynamical space-time structure of PhLO. Hence, PhLO
{\it propagate}, they do not move. The spatial structure is of composite nature
and has two interrelated components represented by the 2-forms $G$ and $G^*$, and
the propagation has two components of constant nature: {\it translational} and
{\it rotational}. The translational component is along isotropic straight
lines in $(\mathbb{R}^4,\eta)$ with constant speed $c$.  The rotational
component of propagation is also of constant nature and follows the
special rotational properties of PhLO's spatial structure.  This
intrinsically consistent dual
nature of PhLO demonstrates itself according to the rule: {\bf no
translation (rotation) is possible without rotation (translation)}.

While the translational component of propagation is easily accounted
through the (arbitrary chosen in general) isotropic autoparallel vector
field $\bar\zeta$, the rotational component of propagation was introduced
in our model making use of two things: the non-integrability properties of
the induced by $\bar\zeta$ two 2-dimensional differential/Pfaff systems,
and the complex structure $\mathcal{J}$.  This approach brought the
following important consequences:

1. It automatically led to the required translational-rotational
consistency.

2. The rotational properties of the solutions are
{\it intrinsic} for the PhLO nature, they are transversal to $\bar\zeta$,
they are in accordance with the action of the complex structure
$\mathcal{J}$ on the 2-forms $G$ and $G^*$,  and they allow the
characteristics {\it amplitude} $\phi$ and {\it phase} $\psi$ of a
solution to be correctly introduced.  Moreover, the rotation is of
periodical nature, and a helical spatial structure along the spatial
direction of propagation is allowed, so, the corresponding rotational
properties differ from those in the case of rotation of a solid as a whole
around a point or axis.

3. The spatial shape and translational properties of a solution are carried by
the amplitude $\phi :\ L_{\bar\zeta}\phi=0$, and the rotational ones are
carried by the phase $\psi$: $L_{\bar\zeta}\psi\neq 0$.

4. The curvature, considered as a measure of the available
non-integrability, or equivalently as a $(\bar{A^*},\bar{\zeta})$-directed
strain $D(\bar{A^*},\bar{\zeta})$
(or as a $(\bar{A},\bar{\zeta})$-directed strain $D^*(\bar{A},\bar{\zeta})$)
, is non-zero only if the phase $\psi $
is NOT a running wave along $\bar\zeta:$\ $L_{\bar\zeta}\psi\neq 0 $,
hence, roughly speaking, curvature means rotation and vice versa.

5. Quantitatively, the curvature $\mathbf{R}$ is {\it obtained} to be
proportional to the product of the energy-density $\phi^2$ and the phase change
$L_{\bar{\zeta}}\psi=const$ along $\bar\zeta$ of a solution
: $\mathbf{R}=l_o\,\phi^2$. We recall that in General Relativity (GR) a
proportionality relation between the energy-momentum density of
{\it non-gravitational} fields and the correspondingly contracted riemannian
curvature is {\it postulated}, while here it is {\it obtained} in the most
general sense of the concept of curvature, namely, as a measure of Frobenius
non-integrability (of correspondingly defined subdistributions of an integrable
distribution which is meant to represent a PhLO). So, if we believe that
the gravitational field carries energy-momentum, this result could be
considered as a strong support for the GR ideology if the gravitational
energy-momentum is manifestly included and treated in the same way as
the energy-momentum characteristics of all other fields.

6. A natural {\it integral} measure of the rotational properties of a
solution appears to be the product $ET$, i.e. the action for one period
$T= 4l_o/c$, which is in accordance with the Planck formula $ET=h$.

Together with the allowed finite nature of the solutions these properties
suggest the following understanding of the PhLO's time-stable dynamical
structure:  {\bf PhLO MUST always propagate in a translational-rotational
manner as fast as needed in order to "survive", i.e.  to overcome the
instability (the destroying tendencies), generated by the available
non-integrability}. In other words, every free PhLO has to be able to
supply immediately itself with those existence needs that are constantly
put under the non-integrability destroying influence. Some initial steps
to understand quantitatively this "smart" nature of PhLO in the terms used
in the paper could be the following.

Recalling the two 2-forms $G=A\wedge\zeta$ and $G^*=A^*\wedge\zeta$ we see
that when $L_{\bar\zeta}\phi^2=0$ then the corresponding subsystems keep the
energy-momentum carried by each of them:
$i(\bar{G})\mathbf{d}G=i(\bar{G}^*)\mathbf{d}G^*=0$. On the other hand the
relation
$i(\bar{G})\mathbf{d}G^*=-i(\bar{G}^*)\mathbf{d}G=-\varepsilon\mathbf{R}.\zeta $ may
be physically interpreted in two ways. FIRST, differentially, $G$
transfers to $G^*$ so much energy-momentum as $G^*$ transfers back to $G$,
which goes along with the previous relations stating that $G$ and $G^*$
keep their energy-momentum densities. Each of these two quantities
$i(\bar{G})\mathbf{d}G^*$ and $i(\bar{G}^*)\mathbf{d}G$ is equal (up to a
sign) to $\mathbf{R}.\zeta$, so, such mutual exchange of energy-momentum
is possible only if the non-integrability of each of the two Pfaff
2-dimensional systems $(A,\zeta)$ and $(A^*,\zeta)$ is present, i.e. when
the curvature $\mathbf{R}$ is NOT zero and is measured by the same
non-zero quantity.

Since the curvature implies outside directed flow (with respect to the
corresponding 2-dimensional distribution) this suggests the SECOND
interpretation: the energy-momentum that PhLO might lose differentially in
{\it whatever way} by means of $G$ is differentially and simultaneously
supplied by means of $G^*$, and vice versa. We could say that every PhLO
has two functioning subsystems, $G$ and $G^*$, such, that the
energy-momentum loss through the subsystem $G$ generated by the
nonintegrability of $(A,\zeta)$, is gained (or supplied) back by the
subsystem $G^*$, and vice versa, and in doing this PhLO make use of the
corresponding rotational component of propagation supported by appropriate
spatial structure. All this is mathematically guaranteed by the isotropic
character of $G$ and $G^*$, i.e.  by the zero values of the two invariants
$G_{\mu\nu}G^{\mu\nu}=G^*_{\mu\nu}G^{\mu\nu}=0$, and by making use of the
complex structure $\mathcal{J}$ as a rotation generating operator.

Let's now try to express this dually consistent dynamical nature of PhLO
by {\it one} object which satisfies {\it one} relation. We are going to
consider $G$ and $G^*$ as two vector components of a
$(\mathbb{R}^2,J)$-valued 2-form, namely, $\Omega=G\otimes I+G^*\otimes
J$. Applying the exterior derivative we get
$\mathbf{d}\Omega=\mathbf{d}G\otimes I+\mathbf{d}G^*\otimes J$. Consider
now the $(\mathbb{R}^2,J)$-valued 2-vector $\bar\Omega=\bar{G}\otimes
I+\bar{G^*}\otimes J$. The aim we pursue will be achieved through defining
the object ($\vee$ is the symmetrized tensor product) \[
(\vee,i)(\bar\Omega,\mathbf{d}\Omega)\overset{\text{\small def}}{=}
i(\bar{G})\mathbf{d}G\otimes I\vee I + i(\bar{G^*})\mathbf{d}G^*\otimes
J\vee J+ \Big[i(\bar{G})\mathbf{d}G^*+i(\bar{G^*})\mathbf{d}G\Big]\otimes
I\vee J \] and put it equal to zero:
$(\vee,i)(\bar\Omega,\mathbf{d}\Omega)=0$.

We note that this last relation $(\vee,i)(\bar\Omega,\mathbf{d}\Omega)=0$
represents the dynamical equations of the vacuum Extended Electrodynamics
(Donev, Tashkova 1995, 2004). In particular, this equation contains all
solutions to the Maxwell vacuum equations, and the solutions obtained in this
paper are a special part of the full subset of nonlinear solutions to these
nonlinear equations.

From a general point of view it deserves to emphasize once again that the
available curvature 2-forms of nonintegrable subdistributions of an integrable
one, represent a natural instrumentarium for describing and understanding the
structure of the complex of internal exchange processes, which
processes guarantee the time-stability of a composit physical object/system
having dynamical structure. The simple case of PhLO considered here suggests
further applications of this approach to physical objects/systems of more
complicated dynamical internal structure.

\vskip 0.3cm

In conclusion, spatially finite field models of
PhLO can be built in terms of complex valued functions and in terms of
isotropic 2-forms on Minkowski space-time, and these two approaches can be
related.  In these both cases substantial role play the complex structures
$J$ and $\mathcal{J}$ as rotation generating operators, carrying
information in this way about the spin properties of PhLO.

The PhLO's longitudinal sizes and rotational orientations can be
determined by the constant parameter combination $\kappa l_o$, and their
effective transversal sizes may be considered to depend on the energy-density
carried by them. The time-stability is guaranteed, on one hand, by the internal
energy-momentum exchange between the two non-integrable 2-dimensional
differential/Pfaff systems, and on the other hand, by a (possible) dynamical
harmony with the outside world.

The significance of the concepts of Frobenius curvature and
$\bar{\zeta}$-directed strain appear to be of primary importance in our
approach.

The results presented in the paper clearly suggest that the Frobenius
integrability theory appears to be important mathematical scheme to be used in
theoretical physics when time-stable physical systems of composite structure,
i.e. considered as built of relatively time-stable mutually interacting
subsystems, are studied. The arizing Frobenius curvatures acquire natural
interpretations of objects used to build the mathematical representatives of
the corresponding energy-momentum exchanges, i.e. force fields. Interaction of
the physical system with the outside world can also be incorporated in this
scheme.
\vskip 0.3cm We kindly acknowledge the support of the Bulgarian
Science Research Fund through Contract $\Phi/1515$.
\vskip 0.3cm
\newpage
{\bf References} \vskip 0.3cm Dainton, J., 2000,  {\it Phil. Trans. R. Soc. Lond.
A}, {\bf 359}, 279 \vskip 0.3cm De Broglie, L. V., 1923,  {\it Ondes et
quanta}, C. R. {\bf 177}, 507 \vskip 0.3cm Donev, S., Tashkova, M., 1995, {\it
Proc. R. Soc. Lond. A} {\bf 450}, 281 (1995);

 arXiv:hep-th/0403244
\vskip 0.3cm
Einstein, A. 1905, {\it Ann. d. Phys.}, {\bf 17}, 132
\vskip 0.3cm
Godbillon, C., 1969,  {\it Geometrie differentielle et mecanique analytiqe},
Hermann, Paris
\vskip 0.3cm Godbole, R. M., 2003, arXiv: hep-th/0311188
\vskip 0.3cm
Lewis, G. N., 1926, {\it Nature}, {\bf 118}, 874
\vskip 0.3cm
Marsden, J., Hughes, T., 1994,
{\it Mathematical foundations of Elasticity},
Prentice Hall 1983;
Reprinted by Dover Publications, 1994
\vskip 0.3cm
Nisius, R., 2001,  arXiv: hep-ex/0110078
\vskip 0.3cm
Planck, M. 1901, {\it Ann. d. Phys.}, {\bf 4}, 553
\vskip 0.3cm
Speziali, P., 1972, Ed. {\it Albert Einstein-Michele Besso
Correspondence} (1903-1955),

Herman, Paris, 453
\vskip 0.3cm
Stumpf, H., Borne, T., 2001, {\it Annales de la Fond. Louis De Broglie},
{\bf 26}, No. {\it special}, 429
\vskip 0.3cm
Synge, J., 1958, {\it Relativity: the special theory}, Amsterdam, North Holland
\end{document}